\documentclass[12pt]{article}
\usepackage[margin=1in]{geometry}
\usepackage[utf8]{inputenc} % allow utf-8 input
\usepackage[T1]{fontenc}    % use 8-bit T1 fonts
\usepackage{setspace}

\RequirePackage{amsthm,amsmath,amsfonts,amssymb}
\RequirePackage[authoryear]{natbib}
\RequirePackage[colorlinks,citecolor=blue,urlcolor=blue]{hyperref}
\RequirePackage{graphicx}
\RequirePackage{mathtools}
\RequirePackage{algorithm}
\RequirePackage{algorithmic}
\RequirePackage{bbm}
\usepackage{authblk}
\RequirePackage{subcaption}
\RequirePackage{caption}
\RequirePackage{booktabs}
\RequirePackage{xcolor}
\RequirePackage{adjustbox}
\RequirePackage{threeparttable}
\RequirePackage{multicol}
\RequirePackage{multirow}
\RequirePackage{cleveref}

\theoremstyle{plain}

\newcommand*{\eg}{\emph{e.g.}{}}
\newcommand*{\ie}{\emph{i.e.}{}}

\renewcommand{\d}{\mathrm{d}}

\newtheorem{theorem}{Theorem}[section]
\newtheorem{lemma}[theorem]{Lemma}
\newtheorem{assumption}{Assumption}

\newtheorem{proposition}{Proposition}
\theoremstyle{remark}
\newtheorem{definition}[theorem]{Definition}

\newtheorem{example}{Example}
\newtheorem{remark}{Remark}
\title{
Hierarchical Probabilistic Conformal Prediction \\ for Distributed Energy Resources Adoption
}

\author[]{Wenbin Zhou}
\author[]{Shixiang Zhu\footnote{Emails: \url{wenbinz2@andrew.cmu.edu} (Wenbin Zhou) and \url{shixianz@andrew.cmu.edu} (Shixiang Zhu)}}
\affil[]{Carnegie Mellon University}

\date{}

\begin{document}

\maketitle

\begin{abstract}
    The rapid growth of distributed energy resources (DERs) presents both opportunities and operational challenges for electric grid management. Accurately predicting DER adoption is critical for proactive infrastructure planning, but the inherent uncertainty and spatial disparity of DER growth complicate traditional forecasting approaches. Moreover, the hierarchical structure of distribution grids demands that predictions satisfy statistical guarantees at both the circuit and substation levels, a non-trivial requirement for reliable decision-making. In this paper, we propose a novel uncertainty quantification framework for DER adoption predictions that ensures validity across hierarchical grid structures. Leveraging a multivariate Hawkes process to model DER adoption dynamics and a tailored split conformal prediction algorithm, we introduce a new non-conformity score that preserves statistical guarantees under aggregation while maintaining prediction efficiency. We establish theoretical validity under mild conditions and, through empirical evaluation on customer-level solar panel installation data from Indianapolis, Indiana, demonstrate that our method consistently outperforms existing baselines in both predictive accuracy and uncertainty calibration.
\end{abstract}

\section{Introduction}

The rapid proliferation of distributed energy resources (DERs), including rooftop solar panels, energy storage systems, and electric vehicles, is transforming the modern energy landscape. In the United States, DER adoption has accelerated dramatically over the past decade. For example, residential solar photovoltaic (PV) capacity grew from approximately $5$ gigawatts in 2015 to over $30$ gigawatts by 2023, and projections suggest that more than one in seven U.S. homes will have rooftop solar by 2030 \citep{seia2023}. Similarly, the deployment of behind-the-meter battery storage has increased nearly fivefold between 2020 and 2023 \citep{woodmackenzie2023}.
These trends illustrate that DERs are no longer peripheral technologies but are becoming integral components of the electricity grid.

While the widespread growth of DERs presents new opportunities for enhancing grid flexibility, resilience, and decarbonization, it also introduces significant operational challenges. A key concern is that excessive DER penetration within localized areas can strain the distribution infrastructure, leading to voltage fluctuations, reverse power flows, and potential damage to substations and transformers \citep{horowitz2019overview, willems2022forecasting}. For instance, circuits that were originally designed for unidirectional power flows can experience overload or protection coordination issues when a high concentration of rooftop solar feeds energy back into the grid \citep{mack2017impacts}. Moreover, the stochastic nature of DER outputs, particularly from solar and wind sources, complicates traditional grid planning and real-time operations \citep{zhang2019uncertainty, mohammed2021management}.

\begin{figure}
    \centering
    \includegraphics[width=1.0\linewidth]{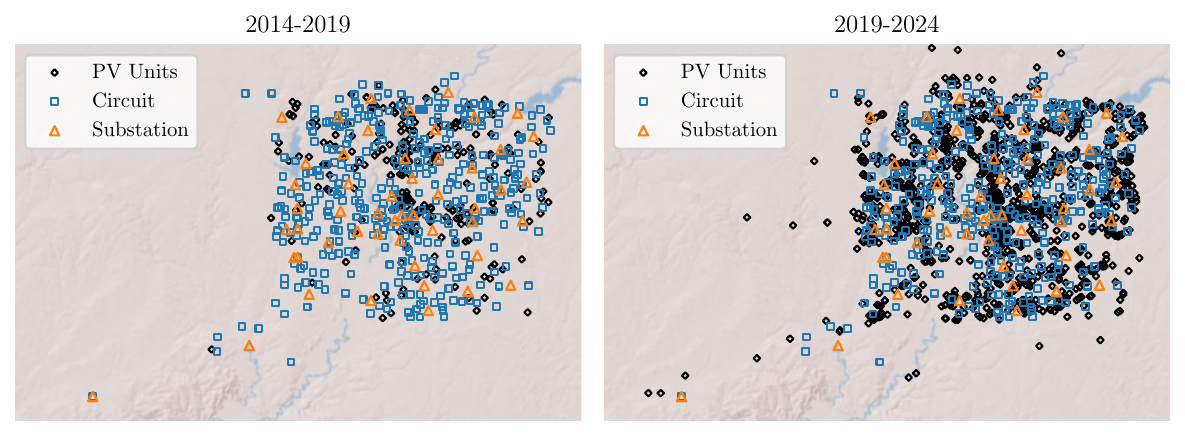}
    \caption{Geographical distribution of incremental photovoltaic (PV) unit installations collected from a real-world dataset. The most recent five-year interval (2019–2024) shows significantly higher growth and greater spatial disparity compared to the preceding interval (2014–2019), highlighting the need for reliable prediction and uncertainty quantification to support future planning and infrastructure development.
    }
    \label{fig:raw-data}
\end{figure}

To manage these challenges, utilities and system operators increasingly rely on predictive analytics to forecast DER growth and proactively plan infrastructure upgrades and operational strategies \citep{khator1997power, georgilakis2015review}. Accurate forecasting enables preemptive actions such as targeted reinforcement of vulnerable circuits \citep{bagheri2018resilient}, optimal siting of energy storage systems \citep{wu2020optimal, novoa2019optimal}, and informed investment decisions for distribution grid modernization. Several utilities have integrated adoption forecasts into their hosting capacity analysis to determine where additional DER interconnections can be accommodated without extensive upgrades \citep{stanfield2021key}.

However, reliably predicting the growth of DERs remains a formidable task. Adoption patterns are highly uncertain over time, influenced by evolving policy incentives, customer preferences, technology costs, and local socio-economic factors. Furthermore, the growth of DERs exhibits pronounced regional disparities: affluent communities often experience faster DER adoption due to higher initial capital affordability and favorable policy environments, while lower-income or rural areas lag behind \citep{barbose2022residential}. These heterogeneous spatio-temporal dynamics, as shown in \Cref{fig:raw-data}, introduce significant volatility and complexity into modeling efforts.
Therefore, rather than relying solely on point predictions that may carry high errors, it is crucial to quantify the uncertainty surrounding DER growth forecasts to enable robust infrastructure planning and operational risk management.

Another key challenge lies in the hierarchical nature of the electric grid. A typical distribution network consists of multiple substations, each connected to numerous circuits. Operational decisions must often be made at both the circuit and substation levels, requiring predictive models to provide coherent forecasts across these scales \citep{hyndman2009optimal, almeida2016hierarchical}. Simply generating independent circuit-level predictions is insufficient: aggregation must preserve statistical guarantees to ensure that uncertainty estimates at the substation level remain valid. For instance, if circuit-level prediction intervals are naively aggregated, the resulting substation-level interval may either under- or overstate the true uncertainty, leading to planning decisions that are either unnecessarily conservative or dangerously optimistic. 

In this paper, we propose a novel distribution-free framework, called Hierarchical Probabilistic Conformal Prediction (\texttt{HPCP}), which constructs prediction intervals that satisfy both circuit-level and substation-level validity. Our method builds upon the split conformal prediction framework \citep{papadopoulos2002inductive, vovk2005algorithmic, lei2013distribution} by partitioning the data into a training set and a calibration set. The training set is used to fit a probabilistic prediction model based on multivariate Hawkes processes \citep{zhu2021quantifying, zhu2022spatiotemporal}, which captures the spatio-temporal dynamics of future DER installations across circuits. The calibration set is then used to compute prediction errors via a carefully designed non-conformity score. Prediction intervals are constructed by estimating the appropriate quantiles of these scores, ensuring that the resulting intervals satisfy the prescribed coverage guarantees at both the circuit and substation levels.
Through empirical evaluation on both synthetic data and real customer-level solar installation data from Indianapolis, Indiana, we demonstrate that our approach consistently outperforms existing predictive and uncertainty quantification baselines. Our results underscore the importance and feasibility of structured uncertainty quantification in supporting reliable and resilient grid planning under the evolving landscape of distributed energy adoption.

The key contribution of our proposed framework is its topology-aware design of the non-conformity score, which enables valid aggregation from circuit-level to substation-level predictions. Specifically, we leverage the grid topology to adjust each circuit’s non-conformity score by incorporating information from ``sibling circuits''---those connected to the same substation. This adjustment strategically aligns the lower and upper bounds of the aggregated predictions, ensuring substation-level validity without substantially sacrificing efficiency. Compared to more conservative or naive aggregation strategies, our approach achieves sharper prediction intervals while rigorously maintaining statistical guarantees.

The remainder of the paper is organized as follows. Section~\ref{sec:literature} reviews related literature and compares our paper with existing works. Section~\ref{sec:setup} defines the hierarchical conformal prediction objective and outlines its key challenges. Section~\ref{sec:method} describes our proposed algorithm, and Section~\ref{sec:theory} provides theoretical analyses of its validity and efficiency. Finally, Section~\ref{sec:experiment} presents empirical evaluations comparing our method with existing baselines for DER adoption modeling on both synthetic and real-world datasets.

\section{Related Works}
\label{sec:literature}

We organize the related work into three categories. First, we review studies focused on predictive models for DER adoption forecasts. Second, we discuss uncertainty quantification methods and approaches applied to DER-related domains. Lastly, we review the methodological development of conformal prediction.

The mainstream models for predicting distributed energy resource (DER) adoption mainly consist of econometric models, time series models, and Bass diffusion models. Econometric models typically use regression techniques to analyze relationships between key drivers and DER adoption \citep{dharshing2017household, bernards2018development, williams2020empirical}. Time series models, in contrast, leverage historical data to capture statistical dependencies for short-term extrapolation \citep{jung2022regional, wang2019photovoltaic}. The Bass diffusion model, widely used for long-term forecasting, simulates the diffusion process of new technologies over time \citep{willems2022forecasting, wang2021diffusion}.
Beyond these traditional approaches, agent-based modeling (ABM) has been used to model adoption at the individual level through discrete choice models and stochastic processes \citep{zhang2016data}, including hazard models based on Cox-type processes \citep{bollinger2012peer} and dynamic nested logit models \citep{bollinger2024valuing}. These approaches are conceptually related to Hawkes process models \citep{hawkes1971spectra}, which explicitly capture self-exciting diffusion behaviors. Hawkes processes have been successfully applied in energy and infrastructure domains such as outage modeling \citep{zhu2021quantifying}, transportation \citep{zhu2021spatio}, epidemiology \citep{dong2023non}, and criminology \citep{zhu2022spatiotemporal}.
Our work leverages the Hawkes process for modeling individual DER adoption events. The key advantage is its ability to capture fine-grained endogenous influences, making it well-suited for settings involving hundreds of individual circuits. In contrast, the other models focus on macro-level exogenous trends, limiting their accuracy at finer resolutions.

Uncertainty quantification goes beyond the point-prediction models described above by characterizing the more informative distributional structure of modeling trajectories.
In the context of DERs, two main approaches have been explored: probabilistic forecasts and interval forecasts. Probabilistic forecasts \citep{van2018probabilistic, li2023probabilistic} model the full distribution of adoption trajectories, often using deep learning techniques \citep{quan2019survey}. In contrast, interval forecasts \citep{shi2017direct, saez2014fuzzy} are less computationally intensive, focusing on estimating upper and lower bounds through methods such as quantile regression \citep{xu2023quantile}.
However, most existing studies focus on forecasting DER generation rather than adoption trajectories. Modeling adoption is significantly more challenging due to its discrete nature and data sparsity, which makes training overly-parameterized models prohibitive. Moreover, many current approaches either lack theoretical guarantees or rely on strong distributional assumptions, limiting their trustworthiness for deployment in high-stakes, risk-averse real-world applications.

Conformal prediction is a distribution-free framework for constructing prediction sets that provide valid coverage guarantees \citep{papadopoulos2002inductive, vovk2005algorithmic, shafer2008tutorial}. Its goal is to achieve the highest possible efficiency while maintaining validity. Prior studies have shown that incorporating distributional information into the non-conformity score can improve the efficiency of the resulting prediction sets \citep{lei2013distribution, lei2014distribution, izbicki2022cd}. This insight has motivated a line of work that integrates probabilistic models as the base predictors within conformal prediction \citep{wang2023probabilistic, zheng2024optimizing}. Recent studies have demonstrated that adopting non-conformity scores with different shapes, such as ellipsoids \citep{sun2023copula, xu2024conformal} or even non-convex forms \citep{tumu2023multi}, can enhance efficiency under certain distributional settings.
Meanwhile, applying conformal prediction to sequential data presents additional challenges due to its reliance on the exchangeability assumption \citep{tibshirani2019conformal, gibbs2021adaptive, barber2023conformal}. Our work builds on a recent line of research whose key idea is to apply quantile regression over the non-conformity scores to anticipate potential distributional shifts in future steps \citep{xu2021conformal, xu2023sequential, xu2024conformal}.

We also note that there are three concurrent efforts on conformal prediction for temporal point process modeling \citep{dheur2024distribution} and hierarchical data structures \citep{principato2024conformal, baheri2025multi} are also closely related to this study. Though our work differs in several key ways:
First, unlike \citet{dheur2024distribution}, which constructs density level sets from a fitted point process model, our approach draws simulated samples from the fitted model. This enables a more flexible formulation tailored to our setting and integrates seamlessly with the hierarchical non-conformity score we propose.
Second, compared to \citet{principato2024conformal}, our work differs fundamentally in its objective---while their approach leverages hierarchical structure to refine predictive intervals, our goal is to ensure that the predictive intervals themselves adhere to the hierarchical structure. Consequently, our solutions are structurally distinct.
Finally, although \citet{baheri2025multi} also addresses hierarchical alignment of prediction intervals, their framework aggregates intervals through set intersections, whereas ours uses summation. This fundamental difference leads to distinct methodological contributions in their work compared to our work.

\section{Problem Setup}
\label{sec:setup}

\begin{figure*}[t]
    \centering
    \includegraphics[width=0.9\linewidth]{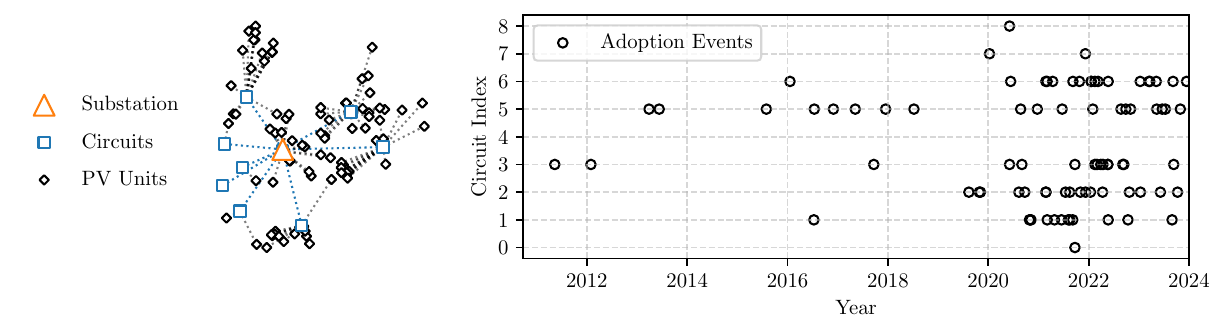}
    \caption{Example illustration of the network topology structure considered in the problem setup.
    \textit{Left}: Affiliated Circuits and DERs (PV Units) of a representative substation.
    \textit{Right}: Adoption event sequences (hollow dots) for eight circuits associated with the substation.
    }
    \label{fig:topo}
\end{figure*}

Consider a utility network consisting of $K$ distribution circuits and $R$ distribution substations. Each substation serves as a ``hub'', connecting and coordinating multiple circuits within the network. The network topology between circuits and substations is defined by a matrix $\mathbf{C} \coloneqq (c_{k,r}) \in \{0, 1\}^{K \times R}$, where $c_{k,r} = 1$ indicates circuit $k$ is associated with substation $r$, and $c_{k,r} = 0$ otherwise. 
Distributed energy resources (DERs), such as rooftop solar panels, are added to the network over time as customers or businesses install them. Each new installation connects to one of these circuits, contributing to local generation or reducing demand. The adoption process is inherently stochastic and varies across circuits and time, influenced by policy, customer behavior, and location \citep{bollinger2012peer, horowitz2019overview}. 
A real example of the network and its DER adoption pattern is provided in \Cref{fig:topo}.
Since $\mathbf{C}$ is inherently a bipartite graph that represents the $0$--$1$ connectivity between circuits and substations within the DER grid topology, we will refer to $\mathbf{C}$ as the (circuit--substation) affiliation matrix throughout this paper following standard network analysis literature terminology \citep{wasserman1994social}.

Our objective is to construct prediction intervals for the number of installations on each circuit within the next time window in the future.
Let $Y_{j,k} \in \mathbb{Z}_+$ denote the number of installations on circuit $k$ during the $j$-th time window of fixed length $\Delta t$ for $j = 1, \dots, n$.
We define the future number of installations on circuit $k$ as $Y_{n+1,k}$, with the vector $Y_{n+1} \coloneqq (Y_{n+1,k})_{k=1}^K$ collecting values across all circuits. The aggregated substation-level installation count is given by $\mathbf{C}^\top Y$. 
We aim to construct lower and upper prediction bounds, $\hat{L} \coloneqq (\hat{L}_k)$ and $\hat{U} \coloneqq (\hat{U}_k)$, such that for a given confidence level $1 - \alpha$, the intervals satisfy:
\begin{align}
    \mathbb{P}\left ( \hat{L}_k \le Y_{n+1,k} \le \hat{U}_k \right ) & \ge 1 - \alpha,\quad k = 1, \ldots, K, \label{eq:val} \\
    \mathbb{P}\left ( [ \mathbf{C}^\top \hat{L}]_r \le [ \mathbf{C}^\top Y_{n+1}]_r \le [ \mathbf{C}^\top \hat{U}]_r \right ) & \ge 1 - \alpha, \quad r = 1, \ldots, R, \label{eq:agg-val}
\end{align}
where $[\cdot]_r$ denotes the $r$-th element of a vector.

The above constraints \eqref{eq:val} and \eqref{eq:agg-val} are directly motivated by operational and management needs in modern power systems with increasing DER adoption. Circuit-level prediction intervals support localized decision-making, enabling targeted actions such as infrastructure upgrades and maintenance planning. However, many critical operations---such as substation-level load management, backup generation deployment, and grid balancing---are conducted at an aggregated level. It is therefore essential that circuit-level intervals collectively yield coherent uncertainty bounds at the substation level. The aggregated constraint \eqref{eq:agg-val} enforces this physical consistency, preventing logical inconsistencies that could arise if aggregated lower and upper bounds were misaligned. 

\begin{figure}[!t]
    \centering
    \includegraphics[width=1.0\linewidth]{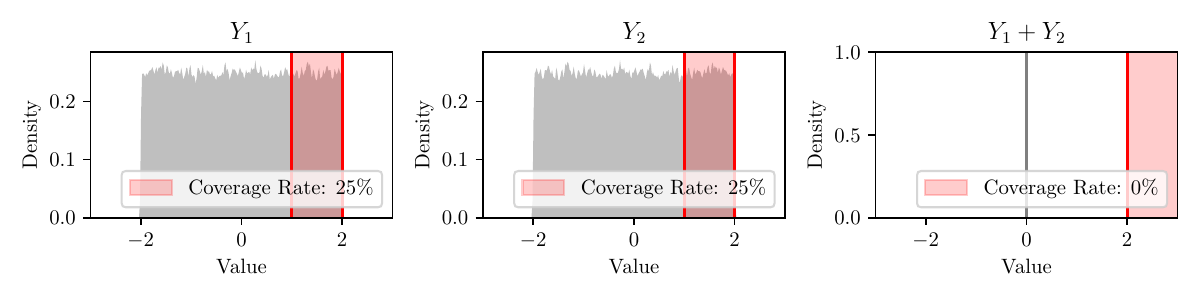}
    \caption{
    Simulation of counterexample provided in Example~\ref{ex} with 10,000 samples, from left to right: $Y_1$, $Y_2$, and $Y_1 + Y_2$.
    The gray area represents the empirical density of the depicted random variable, and the red shaded area represents the constructed prediction interval.
    Note that the density of $Y_1 + Y_2$ is a point mass at the origin in the third plot.
    }
    \label{fig:enter-label}
\end{figure}

We emphasize that satisfying the aggregated constraint \eqref{eq:agg-val} constitutes a non-trivial extension beyond standard conformal prediction, which typically ensures only individual (circuit-level) coverage. In particular, prediction intervals may satisfy \eqref{eq:val} while failing to guarantee substation-level coverage under aggregation. This is illustrated by the following example:

\begin{example}[Toy Example with Two Circuits]
    \label{ex}
    Consider two circuits ($K=2$ and $R=1$) with a target coverage of $1-\alpha = 25\%$, where
    \begin{align*}
    Y_1 \sim \text{Uniform}(-2, 2),\quad Y_2 = -Y_1,
    ~\intertext{with naive prediction intervals:}~
    \hat{L}_1 = \hat{L}_2 = 1, \quad \hat{U}_1 = \hat{U}_2 = 2.
    \end{align*}
    Although each interval $[\hat{L}_i, \hat{U}_i]$ achieves the desired marginal coverage, the aggregate quantity $Y_1 + Y_2 \equiv 0$ falls outside the aggregated interval $[2, 4]$, resulting in a violation of \eqref{eq:agg-val}, as shown in \Cref{fig:enter-label}.
\end{example}

The above example highlights that marginal coverage at the circuit level does not necessarily imply valid coverage at the aggregated substation level, particularly when outcomes are dependent or structurally constrained. In high-dimensional settings, where aggregation is governed by an intricate affiliation matrix $\mathbf{C}$, such inconsistencies become even more subtle and difficult to detect. Thus, developing methods that jointly enforce both circuit-level and substation-level validity is essential for coherent and reliable forecasting in DER planning and grid operations\footnote{While $Y_k$ is continuous in \Cref{ex}, the counterexample can be naturally generalized to discrete settings.}.

\section{Methodology}
\label{sec:method}

This section presents a novel conformal prediction framework for constructing prediction intervals that ensure validity at both the circuit level \eqref{eq:val} and the substation level \eqref{eq:agg-val}.
The framework integrates two main components:
($i$) a multivariate Hawkes process model \citep{reinhart2018review} that captures the spatio-temporal dynamics of distributed energy adoption at the customer level; and
($ii$) a split conformal prediction algorithm \citep{papadopoulos2002inductive} that leverages the fitted model to produce valid and efficient prediction intervals for the total number of future DER adopters at both circuit and substation scales.

\subsection{Preliminaries: Multivariate Temporal Point Processes }
\label{sec:prelim}

Multivariate temporal point processes (MTPPs) \citep{hawkes1971spectra, reinhart2018review} model the occurrence of discrete events over time across multiple dimensions or entities. 
In our context, these events correspond to DER installations over an observed time horizon $T > 0$.
Each event is represented as a tuple $(t, k)$, where $t \in [0, T)$ denotes the installation time and $k \in \{1, \dots, K\}$ indicates the index of the circuit to which the DER is connected. The cumulative number of events up to time $t < T$ is denoted by $N_t$, and the event history up to time $t$ is represented as:
$$
\mathcal{H}_t := \{(t_1, k_1), (t_2, k_2), \dots, (t_{N_t}, k_{N_t})\}.  
$$

Let $\mathbb{N}_k(S)$ denote the counting measure for circuit $k$, representing the number of DER installations that occur over a measurable interval $S \subseteq [0, T)$. For any function $\phi: [0, T) \to \mathbb{R}$, integration with respect to this counting measure is defined as:
$$
\int_S \phi(t)\, d\mathbb{N}_k(t) = 
\sum_{(t_i, k_i) \in \mathcal{H}_T,\; t_i \in S,\; k_i = k} \phi(t_i).
$$

The core quantity in an MTPP is the conditional intensity function $\lambda_k(t \mid \mathcal{H}_t)$, which characterizes the instantaneous rate at which events are expected to occur on circuit $k$ at time $t$, given the event history up to $t$. It is formally defined as:
\begin{equation}
\label{eq:cond-intensity}
    \lambda_k(t \mid \mathcal{H}_t) = \lim_{\delta \to 0} \frac{\mathbb{E}\left[\mathbb{N}_k(t + \delta) - \mathbb{N}_k(t) \mid \mathcal{H}_t \right]}{\delta} = \frac{\d \mathbb{E}[\mathbb{N}_k(t) \mid \mathcal{H}_t]}{\d t}.
\end{equation}

The parameters of a multivariate temporal point process can be learned via maximum likelihood estimation (MLE) using observed event data over the interval $[0, T)$. The log-likelihood of the observed event history $\mathcal{H}_T$ under the model is given by:
\begin{equation}
    \label{eq:ll}
    \ell(\mathcal{H}_T) = \sum_{k=1}^K \left( \int_0^T \log \lambda_k(t \mid \mathcal{H}_t) \, \d \mathbb{N}_k(t) - \int_0^T \lambda_k(t \mid \mathcal{H}_t) \, \d t \right),
\end{equation}
where the first term sums the log-intensity evaluated at each observed event, and the second term penalizes the cumulative expected number of events through the integrated intensity over time. 

\subsection{Probabilistic Prediction via Thinning Sampling}
\label{sec:base}

To model the stochastic process governing DER installations, we adopt a multivariate Hawkes process as defined in \Cref{sec:prelim}, which captures both self-excitation and mutual influence across circuits.
Specifically, we parameterize the conditional intensity function in \eqref{eq:cond-intensity} as follows:
\begin{equation}
    \label{eq:int-param}
    \lambda_k(t \mid \mathcal{H}_t, \mu) = \gamma(t)
    \left[ \mu_k(t) + \sum_{t' < t} \sum_{k'=1}^K \alpha_{k,k'} \cdot \kappa(t, t') \right],
\end{equation}
where $\kappa(t, t') \coloneqq \beta e^{-\beta(t - t')}$ is an exponentially decaying kernel that captures the self- and cross-excitation effects of past DER installations. The parameter $\alpha_{k,k'}$ is an entry of the interaction matrix $\mathbf{A} \in \mathbb{R}^{K \times K}$, quantifying the influence of circuit $k'$ on circuit $k$.
The function $\mu_k(t)$ models the exogenous adoption rate on circuit $k$ as a linear function of external covariates (\eg, socioeconomic and policy features), while $\gamma(t)$ captures the saturation effect---a temporal decay modulating the overall intensity as DER adoption approaches its potential ceiling within the population 
\citep{willems2022forecasting, wang2021diffusion}.

This modeling specification is particularly well-suited to DER adoption for two key reasons:
($i$) \emph{Fine-grained temporal resolution.} It provides a continuous-time framework that models event occurrences at the individual level, thereby avoiding discretization artifacts common in aggregated time series models. This is especially important for avoiding issues such as the modifiable areal unit problem \citep{fotheringham1991modifiable, reinhart2018review}.
($ii$) \emph{Peer and network effects.} The self- and mutually-exciting structure of the intensity function in \eqref{eq:int-param} captures peer influence in DER adoption. The interaction matrix $\mathbf{A}$ and the kernel function $\kappa$ model how prior adoptions by neighboring or connected circuits can increase the likelihood of future installations. This form of contagion-like influence is well-documented in the literature on DER diffusion and social adoption dynamics \citep{bollinger2012peer, gillingham2021social, bollinger2024promotional}.

Given the estimated conditional intensity functions, we generate forecasts of future DER installations over the future interval via probabilistic simulations, which follows a two-stage procedure:
\begin{enumerate}
    \item \emph{Trajectory sampling.} 
    For each circuit $k$, we first simulate a realization of event series over $[t_{j-1}, t_j)$ from the fitted intensity functions $\hat{\lambda}_k(t \mid \mathcal{H}_{t_{j-1}})$ using the thinning algorithm \citep{ogata1981lewis}, which is described in Appendix~A of the Supplementary Material.
    \item \emph{Counting}.
    Define the associated counting measure $\tilde{\mathbb{N}}_k(S)$ for any measurable subset $S \subseteq [t_{j-1}, +\infty)$, representing the number of events occurring on circuit $k$ within $S$. We obtain the sampled realization of the future installation count vector as:
    \begin{equation}
    \label{eq:sampling}
    \tilde{y}_{j} \coloneqq ( \tilde{y}_{j,k} )_{k=1}^K, \quad \text{where} \quad 
    \tilde{y}_{j,k} \coloneqq \tilde{\mathbb{N}}_k([t_{j-1}, t_j)).
    \end{equation}
\end{enumerate}
Repeating this procedure $M$ times yields a collection of conditionally i.i.d. counting vectors $\{ \tilde{y}_j^{(m)} \}_{m=1}^M$ that approximate the predictive distribution of the future DER adoption vector $Y$ under the learned MTPP model $\hat{\lambda}$.

Note that compared to point estimators $\hat{y} = \mathbb{E}[Y]$, the above sampling-based method preserves full information of the predictive distribution, including its variance and tail behavior. This distributional information is crucial for constructing sharp prediction intervals, avoiding undesirable consequences such as underestimation of uncertainty and poorly calibrated intervals that would be incurred by relying solely on point estimates, which is significant in settings where the adoption process is sparse, bursty, or influenced by latent social dynamics.  

\subsection{Hierarchical Probabilistic Conformal Prediction}
\label{sec:alg}

On top of the multivariate Hawkes process model as configured in \Cref{sec:base}, we propose a hierarchical probabilistic conformal prediction algorithm based on the split conformal prediction framework \citep{papadopoulos2002inductive, angelopoulos2021gentle}, adapted to ensure multi-level coverage under structured aggregations. The procedure proceeds in three stages: training, calibration, and prediction. First, the historical event data is partitioned into a training set and a calibration set. The training set is used to fit the multivariate Hawkes process. The calibration set is then used to evaluate the model’s predictive behavior and generate non-conformity scores. Finally, these scores are used to compute quantiles for constructing prediction intervals. The full procedure is summarized in Algorithm~\ref{alg:hcp}.

We begin by splitting the dataset $\mathcal{H}_T$ using a cutoff time index $\tau \in [0, T)$. The data prior to $\tau$, denoted as $\mathcal{D}_{\mathrm{tr}} \coloneqq \{(t_i, k_i) : t_i < \tau\}$, is used to train the multivariate Hawkes process via maximum likelihood estimation using the log-likelihood in Eq.~\eqref{eq:ll}. The post-$\tau$ portion of the data is reorganized into a sequence of calibration pairs $\mathcal{D}_{\mathrm{cal}} = \{(x_j, y_j)\}_{j=1}^n$, where $x_j$ represents the context (history) at time window $j$ for notational simplicity, and $y_j$ denotes the observed adoption count vector (Details of this reorganization are deferred to Appendix~B of the Supplementary Material.)

\begin{algorithm}[t]
\caption{Hierarchical Probabilistic Conformal Prediction}
\label{alg:hcp}
\begin{algorithmic}[1]
    \REQUIRE History events $\mathcal{H}_T$, affiliation matrix $\mathbf{C}$, nominal significance level $\hat \alpha$, sample size $M$, model $\lambda$, prediction horizon $\Delta t$, calibration data size $n$.
    \STATE Initialize $\mathcal{E}_k = \emptyset$ for all $k = 1, \ldots, K$.
    \STATE Split $\mathcal{H}_T$ into training dataset $\mathcal{D}_{\rm tr}$ and calibration dataset $\mathcal{D}_{\rm cal}$;
    \STATE Fit multivariate Hawkes process model $\lambda$ on $\mathcal{D}_{\rm tr}$ by maximizing log-likelihood \eqref{eq:ll};
    \FOR{ $j \in \left\{ 1, \ldots,  n \right\} $}
        \STATE Simulate $M$ outcomes $\tilde{y}_{j}^{(1)}, \ldots, \tilde{y}_{j}^{(M)} \sim \lambda(\cdot | x_j)$;
        \FOR{$k \in \left\{ 1, \ldots, K \right\}$}
            \STATE Use $\mathbf{C}$ to compute $\hat{e}_{k}(x_j, y_j)$ according to  \eqref{eq:score};
            \STATE $\mathcal{E}_k \leftarrow \mathcal{E}_k \cup \{\hat{e}_{k}(x_j, y_j)\}$;
        \ENDFOR
    \ENDFOR
    \STATE Estimate $\alpha$-quantile $\hat{q}_k(\alpha)$ from $\mathcal{E}_k$ for all $k = 1, \ldots, K$;
    \STATE Simulate $K$ outcomes with all observation data $\hat{y}_{n+1}^{(1)}, \ldots, \hat{y}_{n+1}^{(M)} \sim \lambda(\cdot | x_{n+1})$;
    \STATE Return prediction intervals according to \eqref{eq:prediction-interval}.
\end{algorithmic}
\end{algorithm}

For each calibration pair $(x_j, y_j) \in \mathcal{D}_{\mathrm{cal}}$, we simulate $M$ future outcome from the fitted model according to \eqref{eq:sampling}:
\[
\tilde{y}_j^{(1)}, \ldots, \tilde{y}_j^{(M)} \sim \hat{\lambda}(\cdot \mid x_j),
\]
where each sample $\tilde{y}_j^{(m)} \in \mathbb{Z}_+^K$ represents a possible realization of circuit-level adoption counts at time $j$.

Then we define non-conformity scores based on localized residuals. Specifically, we introduce the matrix $\mathbf{S} \coloneqq \mathbf{C} \mathbf{C}^\top \in \{0,1\}^{K \times K}$, where $s_{k,k'} = 1$ indicates that circuits $k$ and $k'$ are associated with the same substation. Let $\mathbf{S}_k$ denote the $k$-th row of $\mathbf{S}$, encoding the ``sibling circuits'' of circuit $k$. The non-conformity score is computed as:
\begin{equation}
    \label{eq:score}
    e_{j,k} = \min_{1 \le m \le M} \left\| \mathbf{S}_k \odot (y_j - \tilde{y}_j^{(m)}) \right\|_\infty,
\end{equation}
where $\odot$ denotes elementwise multiplication, and $\|\cdot\|_\infty$ is the $\ell_\infty$ norm.
This formulation, inspired by probabilistic conformal prediction \citep{wang2023probabilistic}, selects the best-matching simulation among $M$ samples to enhance efficiency while maintaining statistical validity. We highlight that, by focusing only on the residuals over ``sibling circuits'', the score captures local discrepancies that directly impact substation-level aggregates. This stands in contrast to conventional conformal methods, which either treat dimensions independently or pool residuals across all outputs. Our construction leverages the physical network structure to enforce substation-level coverage \eqref{eq:agg-val} more effectively and avoids undue conservativeness.

\begin{figure*}
    \centering
    \includegraphics[width=1\linewidth]{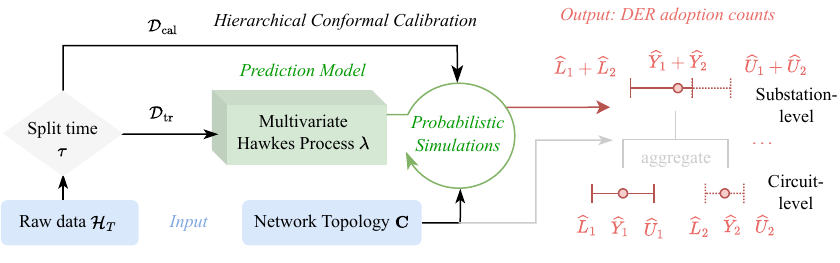}
    \caption{Illustration of our proposed algorithm. The training phase splits the dataset ($\mathcal{D}$), where the training data ($\mathcal{D}_{\rm tr}$) is used to fit the model ($\lambda$), and then used with the calibration data ($\mathcal{D}_{\rm cal}$) to calculate the set of neighbor-aware non-conformity scores ($\mathcal{E}$). Finally, Quantile regression ($\mathcal{Q}$) is fitted over $\mathcal{E}$ to predict the magnitude of the uncertainty and augment the base predictions of $\lambda$ to produce the prediction intervals.
    }
    \label{fig:alg}
\end{figure*}

Given the non-conformity scores and test-time context variable $x_{n+1}$, we simulate $M$ future outcomes, denoted by
$\tilde{y}_{n+1}^{(m)}$,
and construct prediction intervals for each circuit $k$ as:
\begin{equation}
\label{eq:prediction-interval}
\begin{aligned}
\hat{L}_k &= \min_{1 \le m \le M} \tilde{y}^{(m)}_{n+1,k} - \hat{Q}_k(\hat{\alpha}), \quad \forall k = 1, \dots, K,\\
\hat{U}_k &= \max_{1 \le m \le M} \tilde{y}^{(m)}_{n+1,k} + \hat{Q}_k(\hat{\alpha}), \quad \forall k = 1, \dots, K,
\end{aligned}
\end{equation}
where $\hat{Q}_k(\hat{\alpha})$ is the empirical $\hat{\alpha}$-quantile of the calibration non-conformity scores $\{e_{j,k}\}_{j=1}^n$.
To mitigate potential distributional shifts between the calibration and prediction periods, we apply weighted quantile adjustment based on the temporal proximity of each calibration window to the prediction time \citep{xu2023sequential, xu2024conformal}. This enhances robustness when model parameters or underlying adoption dynamics evolve over time. 
An overview of the entire procedure is illustrated in \Cref{fig:alg}.

\section{Theoretical Guarantees}
\label{sec:theory}

In this section, we provide a theoretical analysis of the proposed framework, demonstrating that it satisfies the coverage guarantees in \eqref{eq:val} and \eqref{eq:agg-val} under mild regularity conditions. We also show that, under appropriate hyperparameter choices, the method achieves improved efficiency in the width of prediction intervals. 

\subsection{Coverage Validity}

We begin our validity analysis by introducing a mild assumption on the dependence structure of the residuals.
\begin{assumption}[Strong Mixing of Residuals]
\label{ass}
The residual sequence $\{ y_j - \tilde{y}_j^{(m)} \}_{j=1}^n$ forms a stationary and strongly mixing process with mixing coefficients $\{ w_j \}_{j > 0}$ satisfying
$
\sum_{j > 0} w_j \leq W,
$
where the mixing coefficient is defined as
\[
    w_j \coloneqq \lim_{z \to \infty}\sup_{A \in \mathcal{F}_{0}^{z}, B \in \mathcal{F}_{z+j}^{+\infty}} | \mathbb{P}(A \cap B) - \mathbb{P}(A)\mathbb{P}(B) |.
\]
Here $\mathcal{F}_a^b$ denotes the $\sigma$-field generated by $\{ y_j - \tilde{y}_j^{(m)} \}_{j=a}^b$. 
\end{assumption}
Assumption~\ref{ass} requires that the fitted model $\hat{\lambda}$ captures the temporal patterns of DER adoption sufficiently well, such that the residuals are approximately stationary and weakly dependent. This is a standard condition often imposed in the literature on conformal prediction with dependent data \citep{xu2024conformal}, and does not require model consistency or impose parametric assumptions on the true data-generating process.

Under this assumption, we establish the following finite-sample validity guarantee for our method. The result is stated in terms of the empirical quantile estimator used to construct prediction intervals, though alternative estimators may also be used in practice.
\begin{theorem}[Validity]
\label{thm:val}
Suppose Assumption~\ref{ass} holds, and let $\mathcal{Q}$ denote the empirical quantile estimator. Then, the prediction intervals output by Algorithm~\ref{alg:hcp} satisfy the coverage guarantees in \eqref{eq:val} and \eqref{eq:agg-val}, provided the nominal quantile level is set to
\begin{equation}
    \label{eq:nominal-level}
    \hat{\alpha} \leq \left[ 1 - \frac{1 - \alpha}{1 - \left( {(\log n)^2 W}/{2 n} \right)^{1/3}} - \frac{(W/2)^{1/3}(\log n)^{2/3}}{n^{1/3}} \right]^+,
\end{equation}
where $[\cdot]^+$ denotes the positive part operator.
\end{theorem}

\begin{remark}
    The key technical challenge in proving Theorem~\ref{thm:val} arises from two forms of dependence absent in standard conformal prediction: ($i$) temporal dependence from forecasting future events, and ($ii$) hierarchical dependence due to the aggregation of circuit-level predictions within substations. These dependencies violate the exchangeability assumption typically required for marginal validity, making substation-level coverage non-trivial to guarantee. 
    To address this, the proof relies on two main components: First, the strong mixing condition in Assumption~\ref{ass}, which enables concentration bounds to account for temporal dependence, and second, structural properties of the substation matrix $\mathbf{S}$, including a factorization-to-product property and a weak duality argument. These tools allow us to bound aggregate prediction errors using localized residuals, ultimately yielding an explicit finite-sample bound on the nominal significance level $\hat{\alpha}$ that ensures joint coverage. The detailed proof can be found in Appendix~C of the Supplementary Material.
\end{remark}

Theorem~\ref{thm:val} establishes a finite-sample concentration bound on the empirical quantile estimation error, accounting for residual dependence. 
This result has two practical implications for implementing Algorithm~\ref{alg:hcp}:
($i$) If the mixing coefficient $W$ is known, the nominal level can be explicitly set as the right-hand side of \eqref{eq:nominal-level} to guarantee finite-sample validity;
($ii$) If $W$ is unknown---which is often the case---setting $\hat{\alpha} = \alpha$ yields asymptotic validity as the calibration sample size $n$ grows.

\subsection{Prediction Efficiency}

Next, we establish the efficiency guarantees of our proposed algorithm. We define efficiency as the inverse of the average expected interval size across all circuits, formally given by
$$
V = \frac{1}{K} \sum_{k = 1}^K \mathbb{E} \left[ \hat{U}_k - \hat{L}_k \right],
$$
where a smaller value of $V$ indicates higher efficiency. We assess the efficiency of our method relative to existing state-of-the-art approaches by comparing their respective values of $V$.

This analysis examines two particular ablation variants of \Cref{alg:hcp}:
(\textit{i}) A PCP-style non-conformity score \citep{wang2023probabilistic},
which is equivalent to the score in \eqref{eq:score} when $\mathbf{C}$ is the all-ones vector of length $K$.
(\textit{ii}) A $\ell_p$-norm non-conformity score, which replaces the $\ell_\infty$ norm in \eqref{eq:score} with an $\ell_p$ norm.
These two ablation models represent the predominant existing nonconformity score designs used to address similar objectives as \texttt{HPCP}, ensuring a fair and focused theoretical comparison of the proposed score's ability to balance validity and efficiency.

Let $V_{\rm PCP}$ and $V_p$ denote the efficiency metrics of these two baselines, and let $V_{\rm HPCP}$ denote the efficiency of \Cref{alg:hcp}. We now present the following proposition.
\begin{proposition}[Efficiency]
    \label{cor}
    Suppose that $\mathcal{Q}$ is the empirical quantile estimator used in Algorithm~\ref{alg:hcp}. Then the following efficiency guarantees always hold:
    \begin{align*}
        \text{(a)}\quad & V_{\rm HPCP} \leq V_{\rm PCP}, \\
        \text{(b)}\quad & V_{\rm HPCP} \leq V_p, \quad \forall p \in [0, +\infty).
    \end{align*}
\end{proposition}

\Cref{cor} demonstrates that our proposed non-conformity score achieves higher efficiency than both the PCP-style variant and the $\ell_p$ norm-based alternatives. The intuition behind this result is twofold:
($i$) Compared to the PCP-style variant, our proposed score in \eqref{eq:score} leverages the potentially sparse structure of the affiliation matrix. Given that affiliation matrices are high-dimensional, this exploitation of sparsity reduces redundancy and yields substantially sharper prediction intervals.
($ii$) Compared to $\ell_p$ norm-based scores, our method utilizes the properties of the $\ell_\infty$ norm, ensuring the smallest possible individual prediction interval along each dimension without sacrificing validity. This stands in contrast to traditional multidimensional conformal prediction approaches, where the norm order $p$ is treated as a tunable hyperparameter. The reason lies in the difference of the prediction objective: while prior work seeks to construct a joint conformal prediction region over the entire multidimensional space, our focus is on generating individual prediction intervals for each circuit and substation. 
Additionally, we emphasize that \Cref{cor} does not rely on \Cref{ass}, implying that the efficiency guarantee holds universally across all data-generating mechanisms. This highlights the generality and practical relevance of our result. A formal proof of \Cref{cor} is provided in Appendix~D of the Supplementary Material.

\section{Numerical Studies}
\label{sec:experiment}

In this section, we empirically evaluate the performance of \texttt{HPCP} against ablation variants and baseline models on both synthetic and real data experiments, demonstrating its ability to produce valid and sharp prediction intervals in practice.
The conformal calibration procedure (\ie, nonconformity score) is evaluated with both the synthetic and real data, while the predictive model (\ie, multivariate Hawkes model) is evaluated with real-world data only.
The two parts are complementary in the components they assess and together provide a comprehensive evaluation of the proposed framework.

In both experiments, performance is assessed using two criteria aligned with the previously defined notions of substation-level validity and efficiency, respectively:
($i$) \emph{Coverage rate}: the proportion of substation-level entries for which the prediction interval contains the true DER installation.
($ii$) \emph{Interval size}: the average width of the prediction interval constructed at the circuit level.
Higher coverage rates indicate better validity, while smaller interval sizes indicate greater efficiency. These two metrics are inherently in tension; improving one often comes at the expense of the other. Therefore, it is desirable for a method to strike a good balance between them to ensure practical utility in decision-making, and it is standard in the literature to use both metrics simultaneously to evaluate performance.

It is worth noting that there are two limitations in our numerical analysis:
($i$) We assume that the grid topology network follows a fixed, two-level hierarchy with linear aggregation of circuit-level DER adoption counts.
($ii$) We focus only on the short-term forecasting and uncertainty quantification capabilities of the proposed method.
Therefore, the current empirical study should be interpreted within these modeling assumptions and does not fully capture more complex grid hierarchies or long-horizon planning settings that may arise in real-world DER adoption dynamics. Extending the framework to accommodate richer network structures and longer-term forecasting tasks remains an important direction for future work.

\subsection{Synthetic Data}
\label{sec:syn}

\begin{figure}[!t]
    \centering
    \includegraphics[width=1.0\linewidth]{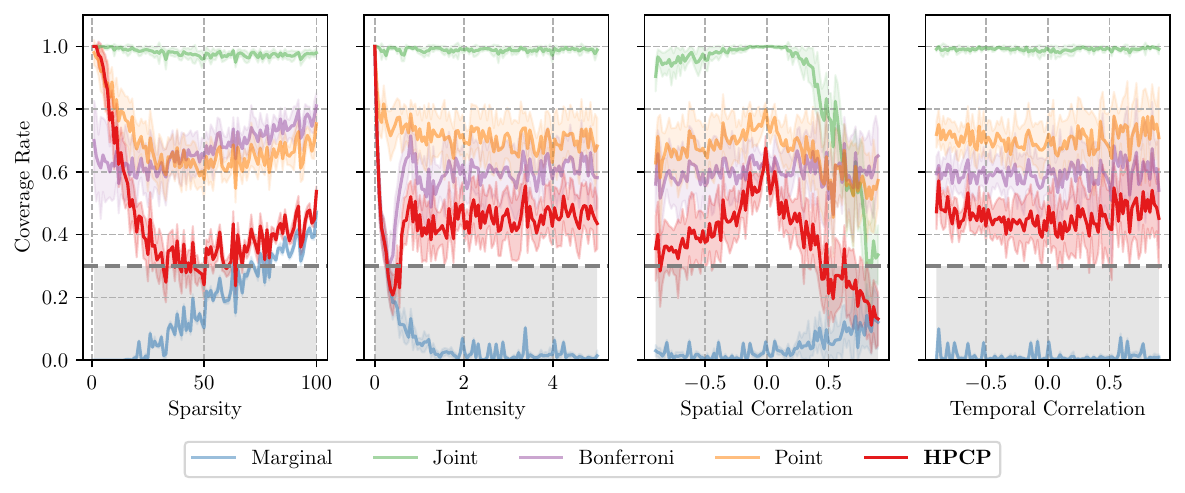}
    \caption{
    Coverage rate of five ablation variants of HPCP by varying four different data generation hyperparameters.
    The gray dashed line represents the reference for the nominal significance level ($\hat \alpha = 70\%$).
    }
    \label{fig:syn-val}
    \includegraphics[width=1.0\linewidth]{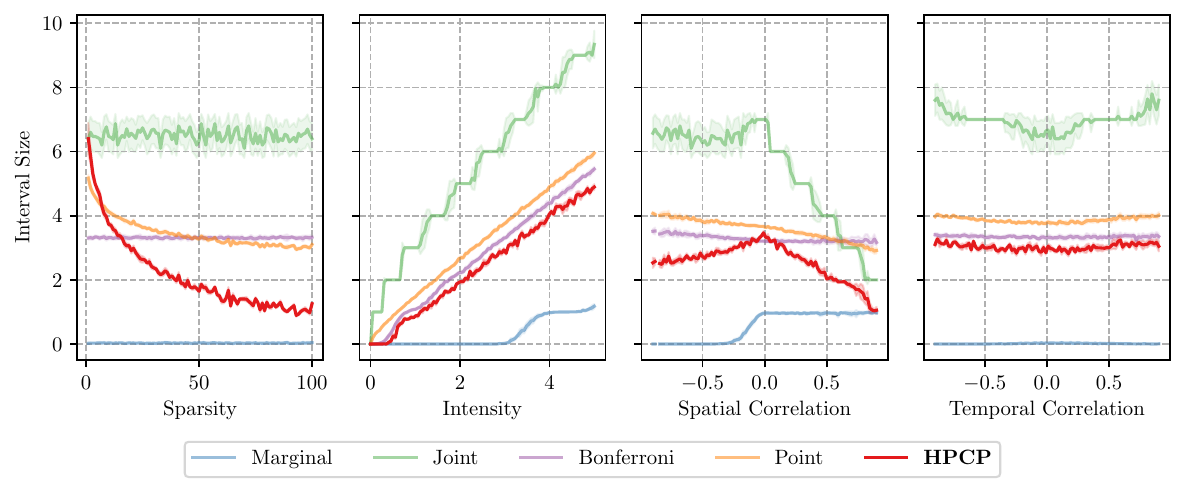}
    \caption{
    Interval size of five ablation variants of \texttt{HPCP} by varying four different data generation hyperparameters.
    }
    \label{fig:syn-eff}
\end{figure}

The synthetic experiments are designed to evaluate the uncertainty quantification ability of the proposed algorithm across four controlled simulation scenarios, in order to assess its applicability to a broader range of real-world settings.
We generate the synthetic data by simulating sequences of intercorrelated Poisson random variables.
Specifically, we apply a Gaussian copula with Poisson marginal transformation to a first-order vector autoregressive (VAR(1)) process.
This approach allows precise control over data properties by tuning the hyperparameters of both the copula and the VAR(1) process.
The key properties we manipulate include:
($i$) Sparsity: The sparsity of the ground-truth circuit-to-substation affiliation matrix, computed as the number of substations. This equivalence is motivated by the intuition that increasing the number of substations leads to a sparser topology under the condition that the number of circuits is fixed.
($ii$)
Intensity: The Poisson parameter value, which is set uniformly across all circuit-level entries.
($iii$)
Spatial correlation: The pairwise rank correlation among different circuit-level entries within the same timestep.
($iv$)
Temporal correlation: The pairwise rank correlation of the same circuit entry across adjacent timesteps.
Unless otherwise specified, all other parameters are held at their default values. 

We introduce four ablation variants of \texttt{HPCP} to be used for comparison: 
($i$) \texttt{Marginal}: Marginal conformal prediction that only guarantees circuit-level validity, equivalent to setting $\mathbf{C}$ in \eqref{eq:score} to a $K \times K$ identity matrix.
($ii$) \texttt{Joint}: Joint conformal prediction, which sets $\mathbf{C}$ in \eqref{eq:score} to an all-ones vector of length $K$, resembling the default setting in many multidimensional conformal prediction algorithms (\eg, PCP \citep{wang2023probabilistic}).
($iii$) \texttt{Bonferroni}: Conformal prediction with Bonferroni correction, which modifies the significance level in \texttt{Marginal} to $\alpha / K$, thereby achieving the same joint validity guarantee as \texttt{Joint}.
($iv$) \texttt{Point}: Conformal prediction using point prediction of the mean as the base prediction. Other specifications are identical to \texttt{HPCP}. 
Across all variants, the predictive model is specified as a constant zero predictor across all settings.
This ensures a fair comparison between different nonconformity score designs.
More implementation details are provided in Appendix~E.1 of the Supplementary Material.

\Cref{fig:syn-val} and \Cref{fig:syn-eff} present the experiment results. 
Across all hyperparameter settings, \texttt{HPCP} consistently achieves empirical coverage rates above the nominal reference line.\footnote{Note that once the nominal coverage rate is exceeded, further increases do not yield additional benefit. Thus, in terms of coverage, our method is not inferior to \texttt{Joint}, \texttt{Bonferroni}, or \texttt{Point}.} Moreover, \texttt{HPCP} attains lower empirical interval sizes compared to all other baselines except \texttt{Marginal}.
In particular, \texttt{HPCP} produces significantly\footnote{Significance is assessed by the non-overlapping standard deviation bands.} smaller interval sizes than \texttt{Point}, indicating that leveraging network topology information and utilizing multiple simulations---as opposed to a single point prediction---enables \texttt{HPCP} to produce prediction intervals that are conservative enough to maintain valid coverage while avoiding excessive width. 

Examining the four individual panels in both \Cref{fig:syn-val} and \Cref{fig:syn-eff}, the following additional insights can be drawn:
($i$) As sparsity increases, \texttt{HPCP} maintains empirical coverage above the nominal level, while achieving a substantial reduction in interval size, significantly smaller than that of both \texttt{Joint} and \texttt{Bonferroni}.
This highlights \texttt{HPCP}’s ability to exploit the sparse structure of the affiliation matrix, particularly in scenarios with many circuits and a moderate number of substations, yielding more informative intervals than traditional methods, consistent with the theoretical result in \Cref{cor}.
($ii$) 
As intensity increases, all methods exhibit larger interval sizes, while their coverage rates remain stable. This reflects conformal prediction’s adaptive nature, where higher intensity (and hence higher variance, in the Poisson setting) leads to appropriately wider intervals to preserve validity.
($iii$) 
Increasing spatial correlation from $0$ to $1$ leads to smaller interval sizes, with coverage remaining above the nominal level. Since DER adoption tends to be spatially correlated \citep{bollinger2012peer}, this suggests that \texttt{HPCP} is well-suited for real-world applications, delivering valid yet compact intervals.
($iv$) 
Changes in temporal correlation have minimal effect on both coverage and interval size. While our theory suggests stronger temporal dependence could affect the validity gap (due to larger mixing coefficients), the empirical results show this effect is negligible, indicating that \texttt{HPCP} remains robust even under temporal dependence where exchangeability may not hold.

\subsection{Real Data}

We evaluate the proposed framework on a real-world dataset of rooftop solar panel installations to demonstrate its practical effectiveness.
The dataset, provided by a local utility, spans from 2010 to 2024, and includes $1,742$ customer-level installation records across $K=245$ circuits and $R=51$ substations, as illustrated in \Cref{fig:raw-data}. Each installation record contains the address, application date, and connectivity details (circuit and substation IDs).
To enhance predictive accuracy, we augment the dataset with five covariates from the utility and the U.S. Census Bureau
\citep{uscensus:portal}: average number of power outages, average electrical load, mean electricity price, average education level, and median household income. 
For predictive model training, we use the proposed multivariate Hawkes process with random initialization. Its parameters are fitted by maximizing log-likelihood using the Adam optimizer with a learning rate of $1 \times 10^{-2}$ for up to $1 \times 10^3$ epochs or until convergence.
All reported metrics in this section are evaluated out-of-sample using a rolling evaluation procedure, where the model is first fitted dynamically over all past time windows of data, and then predicts the outcome value for the next time window, where the time window length is set to be one-month by default.
The prediction result is finally compared with the true observed data in the next time window to compute the evaluated metric. This procedure mimics how predictive analytical procedures are executed in practice, ensuring no data leakage is found in the process, so that the evaluated metric remains informative of how the model would behave in real solar panel adoption deployment. Further visualizations are provided in Appendix E.4.

\begin{table*}[t]
    \caption{
    Comparison of base-scenario mean absolute error (MAE) in an out-of-sample rolling prediction task.
    }
    \centering

    \begin{adjustbox}{max width=\linewidth}
    \begin{threeparttable}
    \begin{tabular}{ c c c c c c c } 
    \toprule[1pt]
    \multirow{2}{*}{\textbf{Method}} & \multicolumn{6}{c}{Time Resolution (Month)} \\
    \cmidrule(lr){2-7}  
    & 1 (Monthly) & 2 & 3 (Seasonal) & 4 & 5 & 6 (Semi-Annual) \\
    
    \midrule
    RNN  & 0.20 $\pm$ 0.03 & 0.29 $\pm$ 0.05 & 0.38 $\pm$ 0.08 & 0.46 $\pm$ 0.13 & 0.55 $\pm$ 0.15 & 0.59 $\pm$ 0.11 \\
    LSTM  & 0.16 $\pm$ 0.03 & 0.26 $\pm$ 0.04 & 0.37 $\pm$ 0.10 & 0.46 $\pm$ 0.10 & 0.54 $\pm$ 0.14 & 0.62 $\pm$ 0.10 \\
    VAR & 0.12 $\pm$ 0.04 & 0.21 $\pm$ 0.06 & 0.29 $\pm$ 0.10 & 0.38 $\pm$ 0.14 & 0.45 $\pm$ 0.17 & 0.53 $\pm$ 0.14 \\
    GP & 0.12 $\pm$ 0.03 & 0.21 $\pm$ 0.06 & 0.30 $\pm$ 0.10 & 0.39 $\pm$ 0.13 & 0.47 $\pm$ 0.17 & 0.54 $\pm$ 0.14 \\
    Discrete Hawkes & 0.08 $\pm$ 0.02 & 0.16 $\pm$ 0.04 & 0.24 $\pm$ 0.07 & 0.32 $\pm$ 0.13 & 0.42 $\pm$ 0.17 & \textbf{0.49 $\pm$ 0.15} \\
    \midrule
    Multivariate Hawkes & \textbf{0.06 $\pm$ 0.02} & \textbf{0.14 $\pm$ 0.04} & \textbf{0.21 $\pm$ 0.08} & \textbf{0.30 $\pm$ 0.14} & \textbf{0.41 $\pm$ 0.19} & \textbf{0.49 $\pm$ 0.16} \\
    \bottomrule[1pt]
    \end{tabular}
    \end{threeparttable}
    \end{adjustbox}

    \vspace{2ex}

    \begin{adjustbox}{max width=\linewidth}
    \begin{threeparttable}
    \begin{tabular}{c c c c c c c c c} 
    \toprule[1pt]
    \multirow{2}{*}{\textbf{Method}} & \multicolumn{6}{c}{Time Resolution (Month)} \\
    \cmidrule(lr){2-7}  
    & 7 & 8 & 9 & 10 & 11 & 12 (Annual) \\
    
    \midrule
    RNN & 0.66 $\pm$ 0.12 & 0.72 $\pm$ 0.13 & 0.76 $\pm$ 0.12 & 0.84 $\pm$ 0.13 & 0.90 $\pm$ 0.12 & 0.93 $\pm$ 0.09 \\
    LSTM & 0.68 $\pm$ 0.11 & 0.76 $\pm$ 0.11 & 0.79 $\pm$ 0.14 & 0.87 $\pm$ 0.13 & 0.89 $\pm$ 0.13 & 0.96 $\pm$ 0.09 \\
    VAR & 0.59 $\pm$ 0.14 & 0.65 $\pm$ 0.15 & 0.73 $\pm$ 0.16 & 0.80 $\pm$ 0.15 & 0.86 $\pm$ 0.13 & 0.92 $\pm$ 0.12 \\
    GP & 0.61 $\pm$ 0.14 & 0.68 $\pm$ 0.16 & 0.75 $\pm$ 0.18 & 0.82 $\pm$ 0.16 & 0.90 $\pm$ 0.15 & 0.94 $\pm$ 0.15 \\
    Discrete Hawkes & \textbf{0.54 $\pm$ 0.13} & 0.63 $\pm$ 0.14 & 0.70 $\pm$ 0.15 & \textbf{0.79 $\pm$ 0.14} & \textbf{0.84 $\pm$ 0.12} & \textbf{0.91 $\pm$ 0.12} \\
    \midrule
    Multivariate Hawkes & \textbf{0.54 $\pm$ 0.15} & \textbf{0.61 $\pm$ 0.16} & \textbf{0.69 $\pm$ 0.18} & 0.81 $\pm$ 0.17 & 0.89 $\pm$ 0.16 & {0.97 $\pm$ 0.15} \\
    \bottomrule[1pt]
    \end{tabular}

    \begin{tablenotes}[flushleft]
        \item \textit{Note:} Numbers are reported as the mean $\pm$ standard deviation across $100$ trials. Results are compared across window sizes from 1 to 12 months (annual) and across the baseline models detailed in the main text.
    \end{tablenotes}
    \label{tab:evaluation}
    \end{threeparttable}
    \end{adjustbox}
\end{table*}

\begin{figure}[!t]
    \centering
    \includegraphics[width=0.9\linewidth]{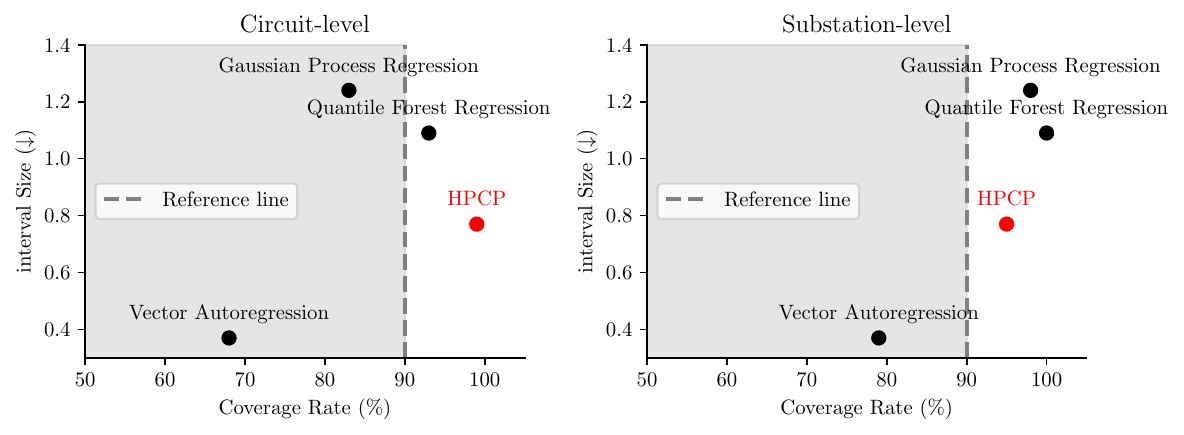}
    \caption{Comparison of uncertainty quantification results for all baselines, \texttt{HPCP} is represented by the red dot. Dots closer to the bottom-right corner represent more desirable methods. The nominal coverage level is set to $90\%$, represented by the gray dashed reference line.}
    \label{fig:uq-comparison}
\end{figure}

We benchmark the performance of \texttt{HPCP} against four competitive statistical and machine learning baselines commonly used in practical DER adoption and load forecasting tasks, including:
($i$) Vector Autoregression (\texttt{VAR}) 
\citep{jung2022regional} is a linear model that jointly captures multivariate temporal dynamics across all locations, with the lag order automatically selected based on the Akaike Information Criterion (AIC).
($ii$) Recurrent Neural Network (RNN) and Long Short-Term Memory (\texttt{LSTM}) 
\citep{wang2019photovoltaic}  are specialized neural network architectures designed for time series modeling. In our implementation, we use a single hidden layer with a hidden dimension of $512$.
($iii$) Gaussian Process Regression (\texttt{GPR}) 
\citep{van2018probabilistic} with a composite kernel function defined as the product of a constant kernel and a Matérn kernel with smoothness parameter $\nu = 0.5$. Specifically, the kernel takes the form
$
\mathbf{K}(x, x') = \sigma^2 \cdot \text{Mat\'ern}_{\nu=0.5}(\|x - x'\|; u),
$
where $\sigma^2 \sim \mathcal{U}(10^{-3}, 10^2)$ is the variance parameter governed by a constant kernel $C(\cdot)$, and $u \sim \mathcal{U}(10^{-5}, 10^2)$ denotes the length scale parameter of the Matérn kernel.
($iv$) Quantile Regression (\texttt{QR})
\citep{lauret2017probabilistic}. We implement a per-coordinate linear quantile regression baseline using a sliding window of lag $L=3$. For each variable $i$, three independent models are fit to predict the $\alpha/2$, $0.5$, and $1 - \alpha/2$ quantiles of $Y_{t, i}$ based on its past $L$ values.
We assume a Gaussian error distribution for both ($i$) and ($ii$) to construct prediction intervals, while ($iii$) and ($iv$) naturally come with uncertainty quantification abilities.
Further implementation details are provided in Appendix~E.2 of the Supplementary Material. 

\Cref{tab:evaluation} summarizes the out-of-sample predictive performance results of all methods, with our proposed model labeled as the Multivariate Hawkes, as defined in \eqref{eq:int-param}.
Standard deviations are calculated over a rolling prediction horizon from 2019 to 2023, and the lowest average errors for each time resolution are highlighted in bold.
Across the $12$ different time resolutions evaluated
(ranging from monthly to annual),
our model achieves the lowest average prediction error in nine cases (from monthly to $9$-monthly). For the remaining three resolutions (from $9$-monthly to annual), its discretized ablation variant (Discrete Hawkes) attains the best performance.
In contrast, machine learning models such as \texttt{RNN} and \texttt{LSTM}, despite their strong expressive power, perform poorly across all experiments.
This underperformance is mainly attributed to their inability to capture the dynamic structure underlying the stochastic nature of DER adoption, as well as their limited capacity to interpolate effectively in data-scarce settings. 
In contrast, the multivariate Hawkes process model achieves strong performance due to its tailored design, which explicitly models excitation, inhibition, and spillover effects. This structure not only captures the underlying stochasticity effectively but also enables stable estimation even with relatively few data samples.
These results indicate the multivariate Hawkes process as a strong predictive foundation for the downstream conformal prediction task, which we detail next.

\Cref{fig:uq-comparison} shows the out-of-sample uncertainty quantification performance of each method at both the circuit and substation levels. The dot representing \texttt{HPCP} lies near the nominal coverage reference line and is positioned closest to the bottom-right corner in both plots. In contrast, other methods either fall to the left of the reference line---indicating insufficient coverage---or produce excessively wide prediction intervals.
These results demonstrate that \texttt{HPCP} not only satisfies the required validity guarantees at both levels but also achieves strong efficiency in interval width. This balance is particularly notable given the stochastic, high-dimensional, and data-scarce nature of the real-world setting, where other generic methods struggle to perform effectively.

\begin{figure}
    \centering
    \includegraphics[width=1.0\linewidth]{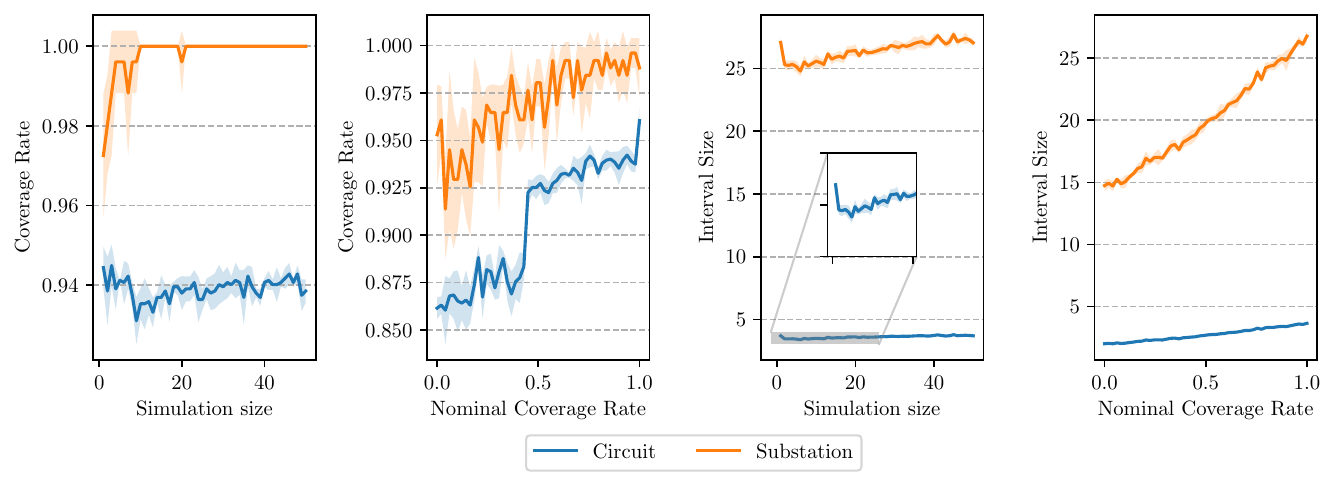}
    \caption{Different specifications of simulated samples $M$ and their effect on the interval sizes and the validity of prediction intervals obtained from our proposed algorithm. The nominal coverage level is $90\%$, represented by the reference line. The shaded area represents the standard error, which is calculated from 5 repeated trials.
    }
    \label{fig:real-param}
\end{figure}

\Cref{fig:real-param} presents an out-of-sample sensitivity analysis of \texttt{HPCP}, showing how its validity and efficiency vary with key hyperparameters: the simulation size $M$ and the nominal miscoverage rate $\hat{\alpha}$. As $M$ increases, the empirical coverage remains stable, while the interval size exhibits a convex trend, reaching a minimum near $M = 5$. This suggests that moderate simulation sizes yield the most efficient intervals. We note that this finding contrasts with the results in \cite{wang2023probabilistic}, which reports monotonically improved efficiency with more samples. 
The difference stems from our conservative interval construction: instead of centering intervals on individual samples, we take the extremal endpoints across all simulations, resulting in slightly wider but more reliable intervals.
For the nominal miscoverage rate $\hat{\alpha}$, the observed coverage consistently exceeds the target $1 - \hat{\alpha}$, while interval size increases roughly linearly as $\hat{\alpha}$ decreases. This demonstrates that \texttt{HPCP} maintains strong validity and adapts across different levels of risk tolerance, justifying its robustness and practicality.

\begin{figure}[!t]
    \centering
    \includegraphics[width = 1\linewidth]{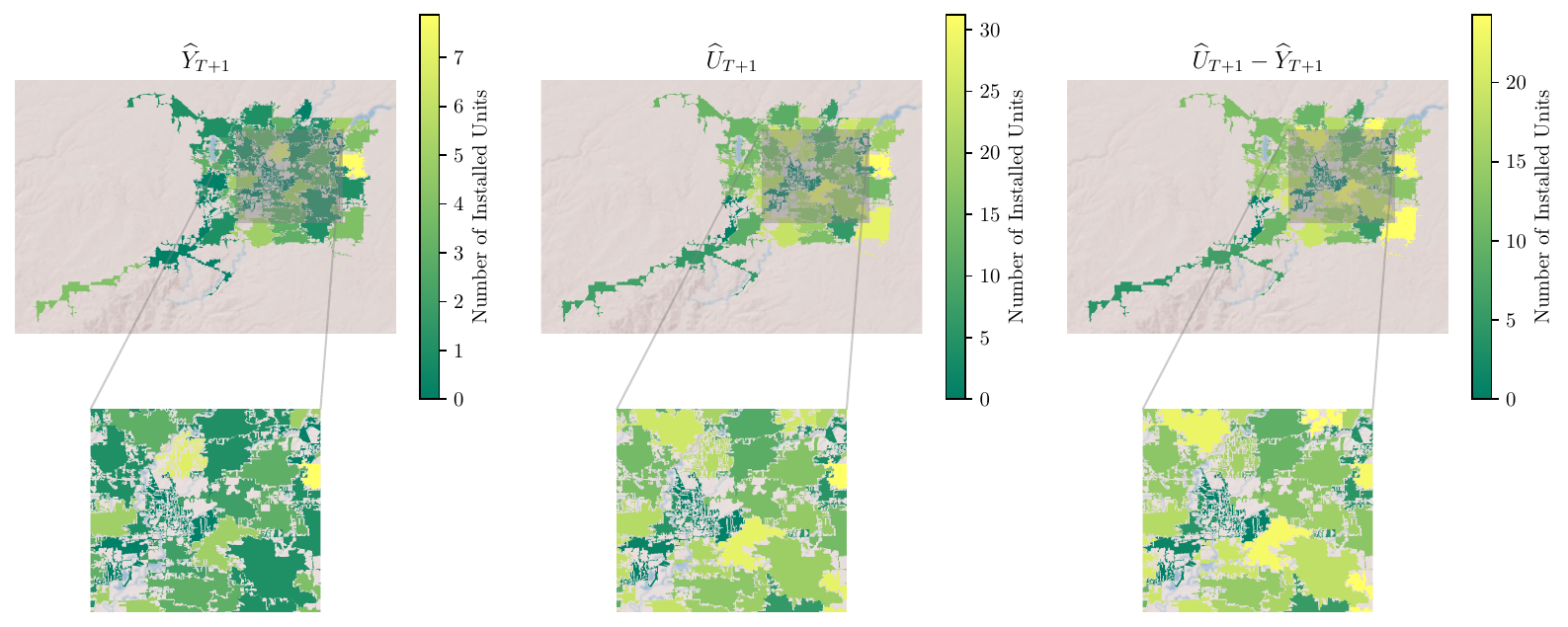}
    \caption{Spatial view of our model's one-year-ahead (2025 - 2026) prediction results at the substation level. From left to right: Base prediction, high-case prediction, and spatial uncertainty map are computed by subtracting the high-case adoption from the base prediction. Each substation is represented by a colored polygon, where the depth of color represents the number of predicted adoption counts.
    }
    \label{fig:spatial-predict}
\end{figure}

\Cref{fig:spatial-predict} provides a spatial view of the one-year-ahead forecast for 2025–2026. The results reveal significant regional disparities in both projected growth and associated uncertainty. In particular, the northern and southwestern regions exhibit strong potential for solar panel adoption. 
The uncertainty map (third plot) mirrors the predicted adoption patterns, indicating that areas with high expected growth also face greater uncertainty, which poses compounded risks for planning decisions. 
These intuitive visualizations facilitate effective communication with stakeholders and can inform more targeted, data-driven planning strategies.

\section{Conclusion}

This paper proposed a novel hierarchical conformal prediction algorithm, referred to as \texttt{HPCP}, to construct prediction intervals for the adoption of distributed energy resources (DER) under the hierarchical constraints inherent to power system settings. By incorporating the network topology connectivity matrix into a new multidimensional nonconformity score, the algorithm effectively calibrates circuit-level uncertainties to ensure validity even when aggregated to the substation level.
We provided theoretical guarantees showing that validity is maintained by slightly inflating the nominal miscoverage rate, and demonstrated that \texttt{HPCP} achieves greater efficiency compared to two common baseline approaches. Experimental evaluations on both synthetic and real-world datasets further confirmed the method’s superior empirical validity and efficiency across diverse data-generating processes and algorithm parameter settings.
In summary, \texttt{HPCP} offers a theoretically sound and practically reliable data-driven tool for topology-aware DER operation planning, particularly beneficial in risk-averse contexts and scenarios with limited adoption data.

The proposed framework has three limitations:
($i$) The current algorithm assumes a two-level hierarchy with linear aggregation of circuit-level DER adoption counts. In practice, grid-level prediction tasks may involve deeper hierarchies (\eg, pole–circuit–substation–region) and nonlinear aggregation functions, such as sigmoid-like relationships that arise when modeling the impact of DERs on grid voltage stability.
($ii$) The algorithm is primarily suited for short-term forecasts due to its data-driven nature, which limits its ability to anticipate events not reflected in historical data, such as the effects of DER-related policies scheduled for future implementation. This constraint makes long-term predictions challenging, particularly in the highly nonstationary and policy-sensitive context of DER adoption.
($iii$) While the assumption of a known and fixed hierarchical structure is justified for the renewable energy adoption setting considered in this paper (See Appendix F for a detailed discussion), it may be restrictive in broader application settings where the networks could be frequently changing or even unknown, such as in healthcare \cite{guha2018emergence}, economics \cite{kouvelis2006supply}, and social analysis \cite{wasserman1994social}.
Future work may enhance the applicability of the proposed algorithm to these settings by considering developing a new nonconformity score that is compatible with non-static and estimated hierarchical structures to ensure validity after aggregation.

\section{Acknowledgments}

We thank Indianapolis Power \& Light Co. dba AES Indiana for funding this project and providing the data and valuable insights that greatly contributed to shaping this research. 
The project was supported by the National Science Foundation (Grant No. CAIG-2425888), and part of this research was performed while the authors were visiting the Institute for Mathematical and Statistical Innovation (IMSI) for a workshop on ``The Architecture of Green Energy Systems'' in 2024, which was co-organized by L.D.A. and supported by the National Science Foundation (Grant No. DMS-1929348).
We also thank Feng Qiu and Xuan Wu for their helpful discussions.

% \begin{funding}
% ...
% \end{funding}

% \begin{supplement}
% \stitle{Proofs and other technical details}
% \sdescription{Appendices containing proofs of all theoretical results and other technical details of the algorithm and numerical experiments that were omitted in the main paper.}
% \end{supplement}

% \begin{supplement}
% \stitle{Python code}
% \sdescription{All experiment codes are provided in Jupyter notebook files named according to their experiment numbers, stored directly in the main folder. The folder ``model'' contains codes for all the baselines (including the proposed algorithm in the paper).}
% \end{supplement}

\bibliographystyle{plainnat}
\bibliography{ref}

\newpage

\appendix

\section{Thinning algorithm}
\label{app:thinning}

\begin{algorithm}[!b]
    \caption{Efficient thinning algorithm for simulating point process}
    \label{alg:thinning}
    \begin{algorithmic}[1]
    \REQUIRE Parameters $\theta$, historic data $\mathcal{H}_0$, forecast horizon $\Delta T$, mark space $\mathcal{I}$;
    \STATE Initialize $\mathcal{H}_t = \mathcal{H}_T$, $t = 0$, $i \sim {\tt uniform}( \mathcal{I})$;
    \WHILE{$t < \Delta T$}
        \STATE Sample $i' \sim {\tt uniform}(\mathcal{I})$; $u \sim {\tt uniform}(0, 1)$; $D \sim {\tt uniform}(0, 1)$;
        \STATE $x' \leftarrow (t, i')$; $\bar{\lambda} \leftarrow \lambda(x' | \mathcal{H}_t)$;
        \STATE $t \leftarrow t - \log u / \bar{\lambda}$;
        \STATE $x \leftarrow (t, i)$; $\tilde{\lambda} \leftarrow \lambda(x | \mathcal{H}_t)$;
        \IF{$D \bar{\lambda} > \tilde{\lambda}$}
            \STATE $\mathcal{H}_t \leftarrow \mathcal{H}_t \cup \{ (t, i)\}$; $m' \leftarrow m$;
        \ENDIF
    \ENDWHILE
    \RETURN A set of continued simulated events $\mathcal{H}_{\Delta T}$, ordered by time.
    \end{algorithmic}
\end{algorithm}

The pseudo-code for the thinning algorithm \citep{ogata1981lewis} is presented in Algorithm~\ref{alg:thinning}.

\section{Details of Data Splitting}
\label{app:split}

\begin{figure}[!b]
    \centering
    \includegraphics[width=0.7\linewidth]{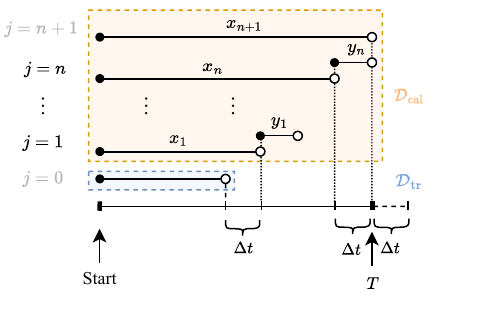}
    \caption{
    Illustration of the calibration data splitting procedure.
    The observed data time horizon $T$ is split into $n$ windows of length $\Delta t$.
    The $j=0$ sequence is used as the training data $\mathcal{D}_{\rm tr}$ (blue box).
    For each $j$-th sequence in $\mathcal{D_{\rm cal}}$ (orange box), events falling in the last time interval of $\Delta t$ are collected as $y_{j}$ and the rest as $x_{\rm j}$ to form a calibration data pair for the point process model.
    Finally, $j=n+1$ represents the sequence of observed events concatenated with the new test events, which is used for final prediction.
    }
    \label{fig:split}
\end{figure}

In this step, we aim to partition our raw dataset, in the form of event sequences, into a training dataset and a calibration dataset.  
Given the prespecified prediction horizon $\Delta t$ and a calibration dataset size $n$ as inputs, we first define a sequence of evenly-spaced temporal grid points $\{ \nu_j \}_{j = 0}^{n}$, where for each $j = 0, \ldots, n$,
\[
\nu_j = T - (n - j) \cdot \Delta t, \quad \forall j = 0, \ldots, n
\]
The training dataset is defined as all events that occur before the zeroth timestep:
$
\mathcal{D}_{\rm tr} = \left\{ (t, k) \in \mathcal{D} : t < \nu_0 \right\}.
$
The rest of the data is reorganized into data pairs and used for the calibration dataset
$
\mathcal{D}_{\rm cal} = \left\{ (x_j, y_j) \right\}_{j = 1}^{n}
$, where for each $j$,
\begin{align*}
    x_j = &~ \{ (t, k) \in \mathcal{D} : t < \nu_{j - 1} \},\\
    y_{k', j} = &~ \left| \left\{ (t, k) \in \mathcal{D} : t \in [\nu_{j - 1}, \nu_j) \wedge k = k' \right\} \right|.
\end{align*}
That is, $x_j$ represents the realization of the historical filtration up to timestep $j$, namely $x_j = \mathcal{H}_{\nu_j}$;
The quantity $y_{k, j}$ represents the number of DER adoption events on circuit $k$ at timestep $j$, which corresponds to the realization of the random variable $Y_{k,j}$ defined in Section~\ref{sec:setup}.
An illustration of the procedure is provided in \Cref{fig:split}.

\section{Proof of Theorem~\ref{thm:val}}
\label{app:proof}

We begin by providing a formal definition of the affiliation matrix and a lemma. They demonstrate some of its useful properties that will be later used in the theorem derivation.

\begin{definition}[affiliation matrix]
    \label{def:topo}
    A matrix $\mathbf{C} \in \left\{ 0, 1 \right\}^{K \times R}$ is a affiliation matrix if
    \begin{equation}
        \label{eq:topo-con}
        | \left\{ r : c_{kr} = 1 \right\} | = 1, \quad \forall k = 1, \ldots, K.
    \end{equation}
    Here, $|\cdot|$ is the cardinality of the set.
\end{definition}

\begin{lemma}[Properties of $\mathbf{S}$]
    \label{lem:S-equiv}
    Denote $\mathbf{S} \coloneqq \mathbf{C} \mathbf{C}^\top$ and $r^*(k) \coloneqq \left\{ r : c_{kr} = 1 \right\}$, then by Definition~\ref{def:topo}:
    \begin{enumerate}
        \item[(a)] For all $k$ and $k'$, if $r^*(k) = r^*(k')$, then $\mathbf{S}_k = \mathbf{S}_{k'}$.
        \item[(b)] For all $k$ and $k'$, $\mathbf{S}_{k,k'} = 1$ if and only if $r^*(k) = r^*(k')$.
    \end{enumerate}
\end{lemma}

\begin{proof}
For (a), observe that for $k'' = 1, \ldots, K$, by definition of $r^*(\cdot)$ and the conditions,
\begin{align*}
    \mathbf{S}_{k,k''}
    & = \sum_{r = 1} c_{kr} c_{k''r} = c_{k r^*(k)} c_{k'' r^*(k)}
    = c_{k' r^*(k')} c_{k'' r^*(k')} = \sum_{r = 1} c_{k'r} c_{k''r} = \mathbf{S}_{k',k''}.
\end{align*}
This means that $\mathbf{S}_k = \mathbf{S}_{k'}$. For (b), following a similar derivation as in (a), there is
$$
\mathbf{S}_{k,k'} = c_{k r^*(k)} c_{k' r^*(k)} = c_{k r^*(k)} c_{k' r^*(k')},
$$
then necessity follows from $r^*(k) = r^*(k')$ and the definition of $r^*(\cdot)$.
To prove sufficiency, using $\mathbf{S}_{k,k'} = 1$ and $\mathbf{S} =  \mathbf{C} \mathbf{C}^\top$, we know that there must exist some $r$ such that $c_{kr} = c_{k'r} = 1$, therefore this $j$ must be equal to both $r^*(k)$ and $r^*(k')$.
\end{proof}

Lemma~\ref{lem:S-equiv}'s two statements jointly describe the property of $\mathbf{S}$ as the substation membership matrix.
We introduce another lemma stating the concentration of the empirical quantile estimator (or more precisely, the empirical CDF) of random variables with temporal dependency outlined in Assumption~\ref{ass}.

\begin{lemma}
    \label{lem:ecdf}
    Under Assumption~\ref{ass}, denote the real marginal CDF and empirical CDF of the residuals as $F$ and $\hat F$, respectively, then there is
    $$
    \mathbb{P}\left(
    \left\| \hat{F} - F \right\|_\infty \leq \frac{(W/2)^{1/3} ( \log n)^{2/3}}{n^{1/3}}
    \right) \geq 1 - \left( \frac{W (\log n)^2}{2 n} \right)^{1/3}.
    $$
\end{lemma}

\begin{proof}
    See the proof of Lemma B.11 of \citep{xu2024conformal}.
\end{proof}

We now present the formal proof for Theorem~\ref{thm:val}.

\begin{proof}[Proof of Theorem~\ref{thm:val}]
    We only establish proof for the substation-level validity \eqref{eq:agg-val}, since the circuit-level validity \eqref{eq:val} is implied by \eqref{eq:agg-val}.
    Recall that the aggregated validity requires that for any $r = 1, \ldots, R$,
    \begin{equation*}
        \mathbb{P}\left( [\mathbf{C}^\top \hat{L}]_{r} \leq [\mathbf{C}^\top Y_{n+1} ]_{r} \leq [\mathbf{C}^\top \hat{U}]_{r} \right) \geq 1 - \alpha.
    \end{equation*}
    We begin the derivation by lower-bounding the left-hand side of the equation:
    \allowdisplaybreaks
    \begin{align*}
        & \mathbb{P}\left( [\mathbf{C}^\top \hat{L}]_{r} \leq [\mathbf{C}^\top Y_{n+1} ]_{r} \leq [\mathbf{C}^\top \hat{U}]_{r} \right) \\
        =~& \mathbb{P}\left( \sum_{k' = 1}^K c_{k'r} \hat{L}_{k'} \leq \sum_{k' = 1}^K c_{k'r} Y_{n+1, k'} \leq \sum_{k' = 1}^K c_{k'r} \hat{U}_{k'} \right) \\
        \geq~& \mathbb{P} \left( \bigcap_{k' : c_{k'r} = 1} \mathbbm{1}\left\{ \hat{L}_{k'} \leq Y_{n+1, k'} \leq \hat{U}_{k'}  \right\} \right), \\
        =~& \mathbb{P} \left( \bigcap_{k' : c_{k'r} = 1} \mathbbm{1}\left\{ \min_{1 \leq m \leq M} | Y_{n+1, k'} - \tilde{y}_{n+1, k'}^{(m)} | \leq \hat{Q}_{k'}(\alpha)  \right\} \right), \\
        =~& \mathbb{P} \left( \bigcap_{k' : c_{k'r} = 1} \bigcup_{1 \leq m \leq M} \left\{ | Y_{n+1, k'} - \tilde{y}_{n+1, k'}^{(m)} | \leq \hat{Q}_{k'}(\alpha)  \right\} \right), \\
        \geq~& \mathbb{P} \left( \max_{k' : c_{k'r} = 1} \min_{1 \leq m \leq M} | Y_{n+1, k'} - \tilde{y}_{n+1, k'}^{(m)} | \leq \min_{k' : c_{k'r} = 1} \hat{Q}_{k'}(\alpha)  \right).
    \end{align*}
    Using the statements and definitions from Lemma~\ref{lem:S-equiv}, for any $k'$ and $k''$ such that $c_{k'r = 1} = c_{k''r = 1} = 1$, then there is $r^*(k') = r^*(k'')$, which by (b) implies that $\mathbf{S}_{k',k''} = 1$. Therefore, continuing on the right-hand side, there is for any $k \in \left\{ k : c_{kr} = 1 \right\}$,
    \begin{align*}
        & \mathbb{P} \left( \max_{k' : c_{k'r} = 1} \min_{1 \leq m \leq M} | Y_{n+1, k'} - \tilde{y}_{n+1, k'}^{(m)} | \leq \min_{k' : c_{k'r} = 1} \hat{Q}_{k'}(\alpha)  \right) \\
        =~& \mathbb{P} \left( \max_{k' : c_{k'r} = 1} \min_{1 \leq m \leq M} \left( \mathbf{S}_{k,k'} \cdot | Y_{n+1, k'} - \tilde{y}_{n+1, k'}^{(m)} | \right) \leq \min_{k' : c_{k'r} = 1} \hat{Q}_{k'}(\alpha)  \right) \\
        \geq~& \mathbb{P} \left( \min_{1 \leq m \leq M} \max_{k' : c_{k'r} = 1} \left( \mathbf{S}_{k,k'} \cdot | Y_{n+1, k'} - \tilde{y}_{n+1, k'}^{(m)} | \right) \leq \min_{k' : c_{k'r} = 1} \hat{Q}_{k'}(\alpha)  \right) \\
        =~& \mathbb{P} \left( \min_{1 \leq m \leq M} \left\|  \mathbf{S}_{k} \odot \left( Y_{n+1, k'} - \tilde{y}_{n+1, k'}^{(m)} \right) \right\|_\infty \leq \min_{k' : c_{k'r} = 1} \hat{Q}_{k'}(\alpha)  \right),
    \end{align*}
    where the first inequality results from weak duality, and the second inequality results from Lemma~\ref{lem:S-equiv} (b).
    By definition of $\hat Q$ is defined as the estimated empirical quantile vector of the non-conformity score \eqref{eq:score}, there is
    $$
    \min_{k' : c_{k'r} = 1} \hat{Q}_{k'}(\alpha) \equiv \hat{Q}_{k}(\alpha).
    $$
    Plugging into the final right-hand side and organizing our derivation, we have now proved that: given any $r \in \left\{ 1, \ldots, J \right\}$ then for any $k \in \left\{ k : c_{kr} = 1 \right\}$, there is
    $$
    \mathbb{P}\left( [\mathbf{C}^\top \hat{L}]_{r} \leq [\mathbf{C}^\top Y_{n+1} ]_{r} \leq [\mathbf{C}^\top \hat{U}]_{r} \right) \geq \mathbb{P} \left( e_k(X_{n+1}, Y_{n+1}) \leq \hat{Q}_{k}(\alpha)\right).
    $$
    By Assumption~\ref{ass}, $\left\{ e_k(X_j, Y_j) \right\}_{j \in \mathbb{Z}}$ is a stationary stochastic process, we denote the stationary distribution as $F_k$ and denote its empirical version computed from the calibration dataset as $\hat F_k$, then there is
    \begin{align*}
        & \mathbb{P} \left( e_k(X_{n+1}, Y_{n+1}) \leq \hat{Q}_{k}(\alpha)  \right) \\
        = ~& \mathbb{P}\left( \hat F_k(e_k(X_{n+1}, Y_{n+1})) \leq 1 - \alpha \right) \\
        = ~&  \mathbb{P}\left( F_k^{-1}(e_k(X_{n+1}, Y_{n+1})) + \hat F_k(e_k(X_{n+1}, Y_{n+1})) - F_k(e_k(X_{n+1}, Y_{n+1})) \leq 1 - \alpha \right) \\
        \geq~ &  \mathbb{P}\left( F_k(e_k(X_{n+1}, Y_{n+1})) \leq 1 - \alpha - \left\| \hat F_k - F_k \right\|_\infty \right) \\
        \geq~ & \mathbb{P}\left( F_k(e_k(X_{n+1}, Y_{n+1})) \leq 1 - \alpha - \left\| \hat F_k - F_k \right\|_\infty \Big| \mathcal{A} \right) \cdot \mathbb{P}(\mathcal{A}),
    \end{align*}
    where $\mathcal{A}$ is defined as the probabilistic event such that for any $k = 1, \ldots, K$,
    $$
    \mathbb{P}(\mathcal{A}) \geq 1 - \left( W (\log n)^2 / 2 n \right)^{1/3}, \quad \mathbb{P} \left( \left\| \hat{F}_k - F_k \right\|_\infty \leq \frac{(W/2)^{1/3} ( \log n)^{2/3}}{n^{1/3}}
    \Bigg| \mathcal{A} \right) = 1.
    $$
    Its existence is guaranteed by Lemma~\ref{lem:ecdf}. Plugging into the above derivation, we get
    \begin{align*}
        & \mathbb{P} \left( e_k(X_{n+1}, Y_{n+1}) \leq \hat{Q}_{k}(\alpha)  \right) \\
        \geq~& \mathbb{P}\left( F_k(e_k(X_{n+1}, Y_{n+1})) \leq 1 - \alpha - \left\| \hat F_k - F_k \right\|_\infty \Big| \mathcal{A} \right) \cdot \mathbb{P}(\mathcal{A}) \\
        \geq~& \mathbb{P}\left( F_k(e_k(X_{n+1}, Y_{n+1})) \leq 1 - \alpha - \frac{(W/2)^{1/3} ( \log n)^{2/3}}{n^{1/3}} \Big| \mathcal{A} \right) \cdot \left(1 - \left( \frac{W (\log n)^2}{2 n} \right)^{1/3} \right) \\
        \geq~& \left( 1 - \alpha - \frac{(W/2)^{1/3} ( \log n)^{2/3}}{n^{1/3}} \right) \cdot \left(1 - \left( \frac{W (\log n)^2}{2 n} \right)^{1/3} \right).
    \end{align*}
    Setting the right-hand side equal to $1 - \alpha$ and substituting the original $\alpha$ notation with the nominal significance level $\hat \alpha$, we get
    \begin{equation*}
        \hat{\alpha} \leq \left[ 1 - \frac{1 - \alpha}{1 - \left( {(\log n)^2 W}/{2 n} \right)^{1/3}} - \frac{(W/2)^{1/3}(\log n)^{2/3}}{n^{1/3}} \right]^+.
    \end{equation*}
    Note that as $n \to \infty$, $\hat \alpha$ eventually converges to $\alpha$. This shows the asymptotic convergence of the algorithm output. Note that the finite-sample bound as in traditional conformal prediction, is typically difficult to derive, as we no longer make the exchangeability assumption \citep{barber2023conformal}.
\end{proof}

\section{Proof for Proposition~\ref{cor}}
\label{app:eff}

\begin{proof}

    Observe that according to the prediction interval construction in~\eqref{eq:prediction-interval}, the expected interval size $V$ can be further expanded as:
    $$
    V = \mathbb{E} \left[ \frac{1}{K} \sum_{k=1}^K \left( \max_{1 \leq m \leq M} \tilde y_{n+1, k}^{(m)} - \min_{1 \leq m \leq M} \tilde y_{n+1, k}^{(m)} \right) \right] + \mathbb{E}\left[ \frac{2}{K} \sum_{k=1}^K \hat Q_k (\alpha) \right].
    $$
    Note the first term is irrelevant to the design of the non-conformity score, thus irrelevant to $p$ and $\mathbf{C}$. Therefore, to prove Proposition~\ref{cor}, we only need to consider the second term, \ie, the value of the estimated quantile. Under the same prespecified significance level $\alpha$, A larger quantile would imply a larger $V$, and vice versa.
    Furthermore, since $\hat Q$ is taken as the empirical quantile estimator, then there is the following property that can be used to compare quantile values:
    \begin{lemma}
        \label{lem:order}
        Let $\{ s_i^{(1)}\}_{i=1}^n$ and $\{ s_i^{(2)} \}_{i=1}^n$ be two sequences of real numbers, and let $\hat Q$ denote the empirical quantile estimator. If $s_i^{(1)} \leq s_i^{(2)}$ for all $i = 1, \ldots, n$, then for any $\alpha \in [0, 1]$, it holds that $\hat Q^{(1)}(\alpha) \leq \hat Q^{(2)}(\alpha)$.
    \end{lemma}

    To prove (a), notice that 
    \begin{align*}
        \min_{1 \leq m \leq M} \left\| \left[ \mathbf{C}^{\top} \mathbf{C} \right]_{k} \odot \left( y_{j,k} - \tilde{y}_{j,k}^{(m)} \right) \right\|_\infty
        & = \min_{1 \leq m \leq M} \max_{1 \leq k \leq K} \sum_{r = 1}^R c_{kr} c_{kr} \cdot  \left( y_{j,k} - \tilde{y}_{j,k}^{(m)} \right) \\
        & \leq \min_{1 \leq m \leq M} \max_{1 \leq k \leq K}  \left( y_{j,k} - \tilde{y}_{j,k}^{(m)} \right) \\
        & = \min_{1 \leq m \leq M} \left\|  y_{j,k} - \tilde{y}_{j,k}^{(m)} \right\|_\infty.
    \end{align*}
    Therefore, by \Cref{lem:order} and the previous argument, there is $V_{\rm HPCP} \leq V_{\rm PCP}$.

    To prove (b), note that by the monotonicity of the $\ell_p$ norm, for any $0 \leq p < \infty$ and any vector $v$, there is $\| v \|_p \geq \| v \|_{\infty}$.
    Therefore, for the nonconformity scores, there is
    $$
    \min_{1 \leq m \leq M} \left\|  \mathbf{S}_{i} \odot \left( y_{j, k} - \tilde{y}_{j, k}^{(m)} \right) \right\|_{p}
    \geq
    \min_{1 \leq m \leq M} \left\|  \mathbf{S}_{i} \odot \left( y_{j, k} - \tilde{y}_{t, k}^{(m)} \right) \right\|_{\infty}.
    $$
    By \Cref{lem:order} and the previous argument, there is $V_{\rm HPCP} \leq V_p$ for any $p \in [0, \infty)$. 

\end{proof}

\section{Details of Numerical Experiments}

\subsection{Synthetic Experiment Configuration}
\label{app:syn}

The vector autoregression sequence is defined as
$$
\mathbf{X}_t = \rho_{\rm temporal} \mathbf{X}_{t -1} + \epsilon_t, \quad \forall t \in \mathbb{Z},
$$
where $\epsilon_t$ is an exogeneous $\mathcal{N}(0, \Sigma)$ variable, and
$$
\epsilon_t \perp \left\{ \mathbf{X}_{t'} \right\}_{t' \leq t}, \quad \Sigma = (1 - \rho_{\rm spatial}) \mathbf{I} + \rho_{\rm spatial} \mathbf{1} \mathbf{1}^\top.
$$
Here, $\rho_{\rm temporal}$ and $\rho_{\rm spatial}$ are hyperparameters, where we have the following proposition:
\begin{proposition}
    \label{prop:cov-structure}
    Let $\mathbf{X}_{t,k}$ be the $k$-th entry of timestep $t$ of the VAR(1) sequence defined above, then the covariance between the random vectors can be derived as:
    \begin{equation*}
        \operatorname{Cov}(\mathbf{X}_{t-1, k}, \mathbf{X}_{t, k}) = \frac{\rho_{\rm temporal} \cdot \rho_{\rm spatial}}{1 - \rho_{\rm temporal}^2}
        ,
        \quad 
        \operatorname{Cov}(\mathbf{X}_{t-1, k}, \mathbf{X}_{t, k}) = \frac{\rho_{\rm spatial}}{1 - \rho_{\rm temporal}^2}
    \end{equation*}
    As a result, the two hyperparameters can be expressed as:
    \begin{align*}
        \rho_{\rm temporal} & = \frac{1 \pm \sqrt{1 + 4 \operatorname{Cov}(\mathbf{X}_{t-1, k}, \mathbf{X}_{t, k})^2}}{2 \operatorname{Cov}(\mathbf{X}_{t-1, k}, \mathbf{X}_{t, k})},
        \\
        \rho_{\rm spatial} & = \operatorname{Cov}(\mathbf{X}_{t-1, k}, \mathbf{X}_{t, k}) \cdot \frac{-1 \pm \sqrt{1 + 4 \operatorname{Cov}(\mathbf{X}_{t-1, k}, \mathbf{X}_{t, k})^2}}{2 \operatorname{Cov}(\mathbf{X}_{t-1, k}, \mathbf{X}_{t, k})^2}.  
    \end{align*}
\end{proposition}

Using \Cref{prop:cov-structure}, one can specify $\rho_{\rm temporal}$ and $\rho_{\rm spatial}$ to generate a sequence with any desired covariance structure.

Next, each entry of $\mathbf{X}_t$ is transformed into a Poisson random variable through a two-step procedure named the Gaussian copula. First, each entry is mapped to a $\text{Uniform}(0, 1)$ variable using the cumulative distribution function (CDF) of the standard normal distribution $\mathcal{N}(0, 1)$. Then, this uniform variable is transformed into a Poisson random variable via the inverse CDF (quantile function) of $\text{Poisson}(\lambda)$, where $\lambda$ is a specified intensity hyperparameter.

Although this transformation may distort the exact covariance structure, it preserves the underlying dependency structure. In particular, monotonic dependence measures such as rank-based correlations (\eg, Spearman’s rho and Kendall’s tau) remain unchanged. As a result, the specified values of $\rho_{\rm temporal}$ and $\rho_{\rm spatial}$ continue to reflect the strength of temporal and spatial dependencies, even though the relationship is no longer linear in terms of covariance or Pearson correlation.
Therefore, in our experiment, we use $\rho_{\rm temporal}$ and $\rho_{\rm spatial}$ as the temporal and spatial correlation hyperparameters, respectively. 

Finally, we note that the proposed multivariate Hawkes process is not used as the predictive model in this setting because:
($i$) This synthetic experiment is designed as an ablation study focusing solely on the nonconformity score design, and incorporating the Hawkes process would introduce unnecessary and uncontrolled confounding factors into the evaluation.
($ii$) Different hyperparameter choices would yield different fitted Hawkes process models, whereas the experiment requires the predictive model to remain fixed across all settings. Adopting a constant zero predictor satisfies this requirement in the simplest and most controlled manner.
The proposed multivariate Hawkes process is evaluated with real data only.

\paragraph*{Sparsity}
We start with the canonical definition of sparsity, which can be expressed as the number of zero entries out of the total number of entries in $\mathbf{C}^*$, 
\begin{equation}
    \label{eq:sparsity}
    \text{Sparsity} = \frac{|c^*_{k,r} = 0|}{K \times R}.
\end{equation}
Then, we have the following proposition, which states how this expression can be simplified to purely depend on the number of substations $R$ based on the definition of $\mathbf{C}^*$.
\begin{proposition}
    \label{prop:sparsity}
    Denote $\mathbf{C}^*$ as the ground-truth affiliation matrix defined in \Cref{def:topo}, suppose sparsity is defined as in \eqref{eq:sparsity}, then
    $$
    \text{Sparsity} = \frac{R - 1}{R}.
    $$
    It can be seen that sparsity is proportional to the number of substations. 
\end{proposition}

\begin{proof}
    Note that starting from the definition of sparsity, we can derive that
    \begin{equation*}
        \frac{|c^*_{k,r} = 0|}{K \times R}
        = 1 - \frac{|c^*_{k,r} = 1|}{K \times R}
        = 1 - \sum_{k = 1}^K \frac{|r: c^*_{k,r} = 1|}{K \times R}
        = 1 - \frac{K}{K \times R}
        = \frac{R - 1}{R}.
    \end{equation*}
    The right-hand side is an increasing function of $R$, which completes the proof.
\end{proof}

\Cref{prop:sparsity} shows that the canonical definition of sparsity scales proportionally with the number of substations. As a result, increasing $R$ leads to a sparser affiliation matrix. Therefore, in our experiments, we use $R$ as the sparsity hyperparameter.

\subsection{Baselines Descriptions}
\label{app:baseline}

The detailed mathematical formulations of the baselines are presented as follows:

\paragraph*{Vector autoregression}
Vector autoregression (VAR) generalizes the univariate autoregressive model to multivariate time series by modeling the current observation as a linear combination of past observations. Specifically, it assumes that the vector-valued response \( Y_t \) depends linearly on a window of previous vectors:
$$
Y_t = \sum_{t' = t - w}^{t - 1} A_{t'} Y_{t'} + \epsilon, \quad \epsilon \sim \mathcal{N}(\mathbf{0}, \boldsymbol{\Sigma}),
$$
where $w$ denotes the window size, $A_{t'}$ are coefficient matrices, and $\epsilon_t$ is a zero-mean Gaussian noise term with covariance matrix $\boldsymbol{\Sigma}$.

\paragraph*{Gaussian Process}
A Gaussian Process (GP) is a non-parametric regression model that assumes the observed data follows a joint multivariate Gaussian distribution. In our setting, we adopt an autoregressive version of the GP. Specifically, let $X_t = (Y_{t - w}, \ldots, Y_{t - 1})^\top$ denote the historical observation vector with window size $w$. Given the training data, the GP assumes that the posterior distribution of $Y_t$ is Gaussian with the following mean and variance:
$$
\mu_t = \mu(x_t) + \mathbf{K}(x_t, \mathbf{x}_{1:t-1}) \mathbf{K}(\mathbf{x}_{1:t-1}, \mathbf{x}_{1:t-1})^{-1}(\mathbf{y} - \boldsymbol{\mu}(\mathbf{x})),
$$
$$
\sigma_t^2 = K(x_t, x_t) - \mathbf{K}(x_t, \mathbf{x}_{1:t-1}) \mathbf{K}(\mathbf{x}_{1:t-1}, \mathbf{x}_{1:t-1})^{-1} \mathbf{K}(\mathbf{x}_{1:t-1}, x_t).
$$
Here, $K : \mathcal{X} \times \mathcal{X} \to \mathbb{R}$ is a kernel function, and $\mathbf{K}$ denotes the Gram matrix evaluated on the inputs.

\paragraph*{RNN and LSTM}
The recurrent neural network (RNN) is a type of artificial neural network designed for modeling sequential data. RNNs have loops (recurrent connections) with the same set of weights applied to the sequential data. Long Short-Term Memory (LSTM) is a special type of RNN that is designed to handle the vanishing gradient problem by using a memory cell and gates to regularize information flow.

\paragraph*{Gaussian Copula}

The copula method \citep{zhang2025weather} estimates a joint multivariate distribution by combining individual marginal distributions with a copula function that captures their dependence structure. A representative choice, used in our experiments, is the \emph{Gaussian copula}.
Given data from a random vector $(X_1, \ldots, X_d)^\top$, we first estimate the marginal distributions $\hat F_i$ (\eg, using empirical CDFs). We then estimate the dependence structure by transforming each variable using the probability integral transform:
$$
Z_i = \Phi^{-1}(\hat F_i(X_i)), \quad i = 1, \ldots, d,
$$
where $\Phi$ is the CDF of the Gaussian distribution, 
and compute the empirical covariance (or correlation) matrix of the transformed vector $ \mathbf{Z} = (Z_1, \ldots, Z_d)^\top$ to obtain $\hat \Sigma$.
The Gaussian copula model generates samples by: ($i$) Drawing $\mathbf{v} \sim \mathcal{N}(\mathbf{0}, \hat \Sigma)$, and then ($ii$) Applying the inverse transform to recover original marginals. The mean prediction under the copula model is defined as:
$$
\mathbb{E}_{\mathbf{v} \sim \mathcal{N}(\mathbf{0}, \hat \Sigma)} \left[ \left( \hat F_1^{-1}(\Phi(v_1)), \ldots, \hat F_d^{-1}(\Phi(v_d)) \right)^\top \right],
$$
which can be approximated via Monte Carlo sampling.
Moreover, uncertainty quantification can be performed by computing entrywise quantiles of the Monte Carlo samples:
$$
\hat Q_{\mathbf{v} \sim \mathcal{N}(\mathbf{0}, \hat \Sigma)} \left[ \left( \hat F_1^{-1}(\Phi(v_1)), \ldots, \hat F_d^{-1}(\Phi(v_d)) \right)^\top ; \alpha \right],
$$
where $\hat Q[\cdot; \alpha]$ denotes the empirical $\alpha$-quantile, applied entrywise. 

\subsection{Covariates Description}

\begin{itemize}
    \item Household median income (2022, inflation-adjusted) in dollars, sourced from the United States Census Bureau data portal, with spatial resolution at the census tract level.
    \item Average total load at the feeder level, provided by the local utility. The raw data consists of individual currents (in amps) for each phase of a three-phase feeder. The load is calculated by summing these currents, multiplying by the system operating voltage, and dividing by the conversion factor ($\sqrt{3}$) to account for the relationship between line-to-line and line-to-neutral voltage. The unit is $\text{kVA}$, and the data spans from the beginning of 2021 to the end of 2023.
    \item System Average Interruption Duration Index (SAIDI) for outages associated with the local utility. SAIDI for a circuit represents the average duration of power interruptions experienced by a customer, calculated as:
    $$
    \text{SAIDI} = \frac{\text{Total minutes of power outages}}{\text{Number of customers}}.
    $$
    The dataset includes the exact coordinates and duration of each outage, as provided by the local utility. The temporal coverage extends from the beginning of 2004 to August 2024.
    \item 
    Total annual average solar energy generation potential across all rooftop areas in the region, measured in kilowatt-hours. The data, sourced from Google Project Sunroof (August 2019), is provided at the census tract level.
\end{itemize}
These covariates are structured as a single vector, with entries corresponding to individual circuits to capture their relative characteristics. This is achieved by averaging the data over its given temporal resolution and then aggregating or disaggregating it to align with the geocoordinates of the circuits.

\begin{figure}[!t]
    \centering
    \includegraphics[width=1.0\linewidth]{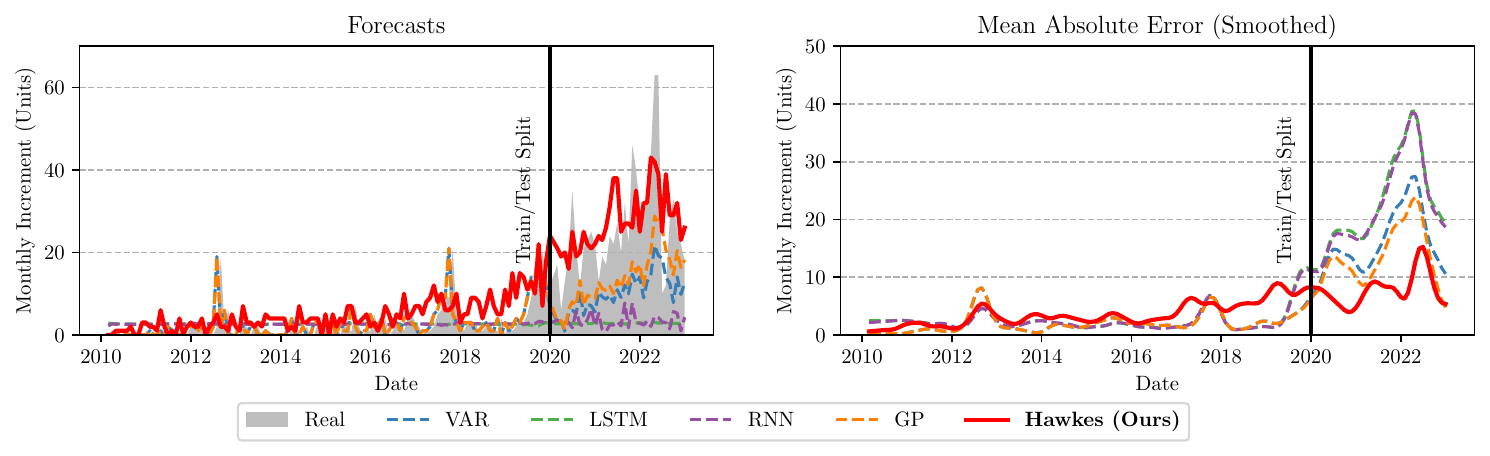}
    \caption{
    \color{black}
    Monthly rolling forecasting results for the entire Indianapolis region.
    The vertical black line indicates the train–test split time point: results to the left are in-sample, while results to the right are out-of-sample.
    Different colors represent different models and baseline methods.
    \textit{Left Panel}: Forecasts.
    \textit{Right Panel}: Mean absolute error between the forecasts and the observed values (gray shaded area), shown after applying a Gaussian smoothing kernel with $\sigma = 2$ to reduce noise and better highlight relative differences in magnitude.
    }
    \label{fig:real_forecast_err}
\end{figure}

\begin{figure}[!t]
    \centering
    \includegraphics[width=1.0\linewidth]{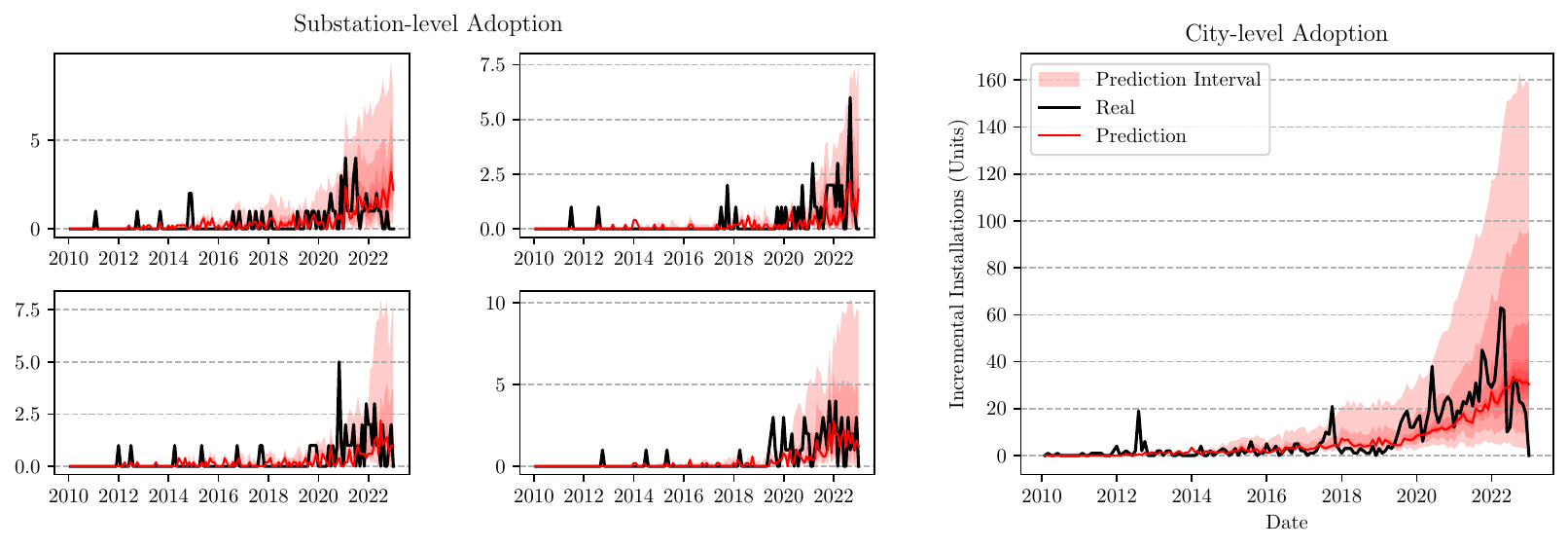}
    \caption{
    \color{black}
    Monthly prediction and prediction intervals of solar panel adoption trajectories. Darker red shade represents a lower miscoverage rate $\alpha \in [0.1, 0.9]$.
    \textit{Left $2 \times 2$ Panel}: Four substation-level solar panel adoption trajectories, corresponding to the four substations with the highest cumulative adoption counts.
    \textit{Right Panel}: Entire Indianapolis region solar panel adoption trajectory, aggregated from all spatial units.
    }
    \label{fig:real_band}
\end{figure}

\begin{figure}[!t]
    \centering
    \includegraphics[width=0.6\linewidth]{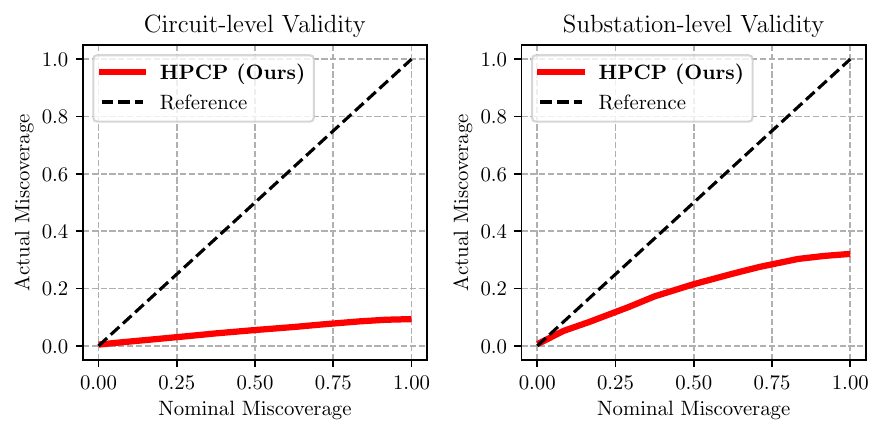}
    \caption{
    \color{black}
    Out-of-sample validity calibration curve, comparing the nominal (prescribed) miscoverage rate $\alpha$ with the actual (empirical) miscoverage rate achieved by HPCP on the real solar panel adoption data.
    A desirable model should trace a curve under the reference line (black dashed line).
    \textit{Left Panel}: Circuit-level validity.
    \textit{Right Panel}: Substation-level validity.
    }
    \label{fig:real_val_tradeoff}
\end{figure}

\subsection{Additional Numerical Results}
To make the real data evaluation results presented in \Cref{sec:experiment} more intuitive and interpretable, we add three figures (\Cref{fig:real_forecast_err}, \Cref{fig:real_band}, and \Cref{fig:real_val_tradeoff}) in the supplementary material, illustrating intermediate evaluation steps.
Specifically, \Cref{fig:real_forecast_err} presents the monthly rolling forecasting results for the entire Indianapolis region. The proposed Hawkes process model in this work achieves superior predictive performance, aligning closely with the observed DER adoption rate and yielding a lower average testing error than other baselines.
This serves as evidence showing that the Hawkes process model forecasts are more accurate than simpler alternatives.
\Cref{fig:real_band} shows the monthly prediction and prediction intervals of solar panel adoption trajectories.
Both the substation-level and the entire Indianapolis-level, the produced rolling forecasts align closely with the real observed data that was not revealed to the model in advance, and produce prediction bands that have a higher empirical coverage rate than their desired coverage rate, as outlined by the validity calibration curve in \Cref{fig:real_val_tradeoff}.
They provide visually straightforward demonstrations of the proposed model's efficacy in real predictive tasks.

\section{Clarifying DER Configuration and Grid Topology}

In this section, we further justify that our assumption on a known and fixed hierarchical grid structure $\mathbf{C}$ is reasonable, what variation factors of the grid are introduced by DER configurations, and whether these factors affect our model's performance.

\paragraph*{Distinction between grid topology and DER configurations}
Grid topology and DER configurations are physically distinct and operationally independent components. In particular, DERs (e.g., solar panels) are not \textit{directly} connected to the grid topology; rather, they are installed at the household level and connected \textit{behind the meter}, supplying power through each household’s metering infrastructure before interfacing with the broader grid. Then, households within a neighborhood (typically on the order of ten) are connected to a common circuit, and circuits are subsequently affiliated with substations.

\paragraph*{What variations do DER configurations introduce?}
DER configurations do not change the grid topology $\mathbf{C}$.
Instead, the main variation they introduce is the change of grid load.
For example, solar panels generate electricity during sunny periods, which can offset a household’s load consumption (e.g., heating and lighting).
More DER adoption in one area can lead to more electricity generation during the daytime, and thus help relieve grid stress in the associated grid infrastructures (i.e., circuits and substations).
Consequently, if higher DER adoption is anticipated in neighborhood $A$ than in $B$, a utility may strategically allocate more infrastructure capacity to the circuits/substations serving $B$ rather than $A$.

This precisely highlights the motivation of our paper---we aim to predict the uncertainty in future DER adoption at different grid hierarchies so that utilities can better anticipate these load variations and make more informed grid planning decisions.
These variations are exactly the objective of our paper.

\paragraph*{Do these variations affect the performance of the method?}
No, variations in the load do not affect the performance of the method. Our model prediction only focuses on the DER adoption pattern rather than the load and its impact on the grid system. The former is a market-driven process that is independent of the latter.
Therefore, as long as assumptions such as exchangeability, stationarity, and the strong mixing condition hold for the DER adoption process, our method enjoys the claimed theoretical guarantees on its performance.

\end{document}